\documentclass[%
 amsmath,amssymb,
 aip,jcp,twocolumn,reprint,noshowpacs
]{revtex4-2}

\usepackage{graphicx}
\usepackage{dcolumn}
\usepackage{bm}
\usepackage{hyperref}

\usepackage{xcolor}

\begin{document}

\title{Pendant Drop Tensiometry: \\ A Machine Learning Approach}

\author{Felix S. Kratz}
\author{Jan Kierfeld}%
 \email{Jan.Kierfeld@tu-dortmund.de}
\affiliation{%
 TU Dortmund University, Germany\\
 Department of Physics
}%

\date{\today}

\begin{abstract}
  Modern pendant drop tensiometry relies on numerical solution
  of the Young-Laplace equation and allow to determine
   the surface tension
   from  a single picture of a pendant drop with high precision.
  Most of these
  techniques solve the Young-Laplace equation many times over to find the
  material parameters that provide a fit to a supplied
  image of a real droplet. Here we introduce
  a machine learning approach to solve
  this problem in a computationally more efficient way.
  We train a deep neural network  to determine the surface tension
   of  a given  droplet shape using a large  training set of  numerically
  generated droplet shapes. We
  show that the deep  learning approach is superior
    to the current state
    of the art shape fitting approach in speed and precision,
    in particular if shapes in the training set reflect the
   sensitivity of the droplet shape with respect to surface tension.
In order to derive such an optimized training set we
    clarify the role of the Worthington number as quality
  indicator in conventional shape fitting and in the machine learning
  approach.
  Our approach demonstrates the capabilities of deep neural networks in
  the material parameter determination from rheological deformation
  experiments in general.
\end{abstract}

\maketitle
\section{Introduction}

Tensiometry is a technique to determine the surface or interfacial
tensions of a fluid interface. Many tensiometry methods are based
on the shape analysis of  liquid drops suspended in air or another liquid.
Available tensiometric methods include the
drop weight method \cite{Tate1864,Harkins1919, Garandet1994, Yildirim2005} and
the oscillating drop method \cite{Fraser1971,Matsumoto2004};
the by far most frequently used tensiometric
technique is the pendant drop method, which is also closely related to
the sessile droplet method as both methods rely on the
shape analysis of a gravity-deformed droplet based on the
Young-Laplace equation. In the pendant drop setup the droplet
typically hangs from the tip of a capillary. Variants
can include, for example additional spherical particles attached
to the droplet \cite{Neeson2014}.
 Pendant liquid drops have been investigated extensively since the
  18th century, however only in the late 20th century numerical solution
  techniques made it possible to extract the surface tension from
  a single picture of a pendant drop with high precision.

\begin{figure}
  \includegraphics[width=0.99\linewidth]{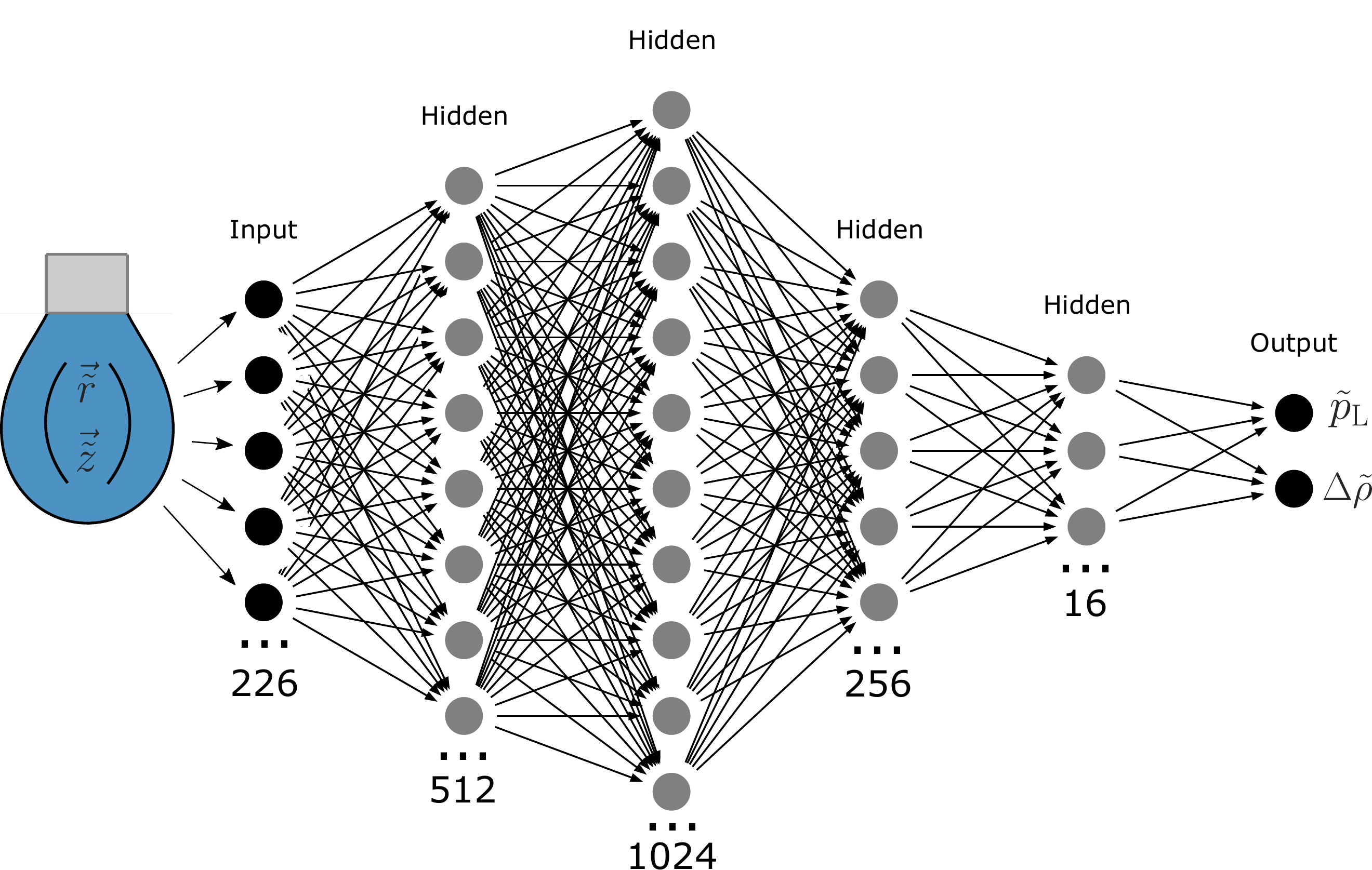}
  \caption{ Visualization of the structure of the
    deep neural network employed for pendant drop tensiometry.
    The  neural network is trained to  solve
    the inverse problem to determine the surface
    tension (which is contained in the dimensionless
      density difference $\Delta\tilde{\rho}$)
    from a pending drop shape.
   }
   \label{fig:network}
 \end{figure}

Before the rise of fast and accessible computer technology the main way to
determine interfacial tension from a pendant drop experiment has been the use
of precomputed tables in which experimentally accessible
dimensionless shape parameters,
such as the ratios  of
the  maximum width $D_E$  of the droplet and
the droplet width $D_S$ a distance $D_E$ from the apex,
are listed with the
corresponding interfacial tension \cite{Andreas1938,Stauffer1965, Juza1997}.

In recent years numerical solution schemes that determine the
interfacial tension from the whole droplet profile became more popular and
viable solution techniques because of
the rapid rise in computers speed
\cite{Hansen1991, Song1996, Ro1997, Dingle2005,  Hoorfar2006, Berry2015}.
Several implementations exist, where only a single
image of a pendant drop and some reference length scale have to be supplied to
get a fully automated fit and a surface tension estimate
\cite{Touhami1996,  Berry2015,  openCapsule}.
At the core of this approach is a numerical
shape fitting scheme
that solves the  Young-Laplace shape equations of the drop many times till
optimal parameters are found, that provide the best match of the
calculated shape to  the supplied image. The
 precision of these methods is often limited  by the resolution of the
supplied image not allowing for a better fit \cite{Berry2015}.

The shape fitting problem is, thus, a classical inverse problem
of finding a material parameter set that minimizes a suitably
defined distance metric between measured  and calculated shape.
In a Bayesian sense we maximize the likelihood of the material
parameters given the measured shape.
The forward problem to calculate a droplet shape
given the surface tension, gravity, pressure, and the diameter
of the capillary can be easily and stably solved  by solving the
shape equations, which are a set of ordinary differential equations
based on the Young-Laplace equation.
The corresponding inverse problem of determining the
surface tension
and pressure given an observed shape is often ill-conditioned  if
the shape becomes insensitive to parameter changes.
In this sense,  pendant drop tensiometry is a paradigm for many
similar inverse problems in rheology.
It has only recently been demonstrated
that machine learning approaches can be useful to  solve such
ill-conditioned inverse problems \cite{Pilozzi2018}.
To our knowledge, an
implementation of a machine learning approach for the
pendant drop problem has never been discussed before and offers a novel
way to think about the general solution of inverse problems in rheology.
So far machine learning applications to  rheological problems
are limited to solving viscoelastic forward
problems with the help of neural networks to replace
full finite element calculations \cite{Kessler2007,DeVries2017}.

The way a deep neural network learns correlations
between input data and output
data is especially helpful if a supervised learning scenario can be
created. For problems in rheology and physics in general this is often the
case, since the forward problem may be sufficiently easy to solve and to
compute, the inverse problem, however, can be exponentially hard to
solve. Generating a large
training set by solving the forward problem many times
and training a deep neural network with this data set to learn the
necessary correlations
to solve the inverse problem can lead to results that even
outperform sophisticated conventional shape fitting
approaches. Additionally, deep neural
networks are lightweight and fast once the network has been trained,
which is essential if high-throughput analysis is required.
We want to explore the capabilities of a machine learning approach to the
inverse Young-Laplace problem as a way to combine the precision of a
forward numerical
solution scheme with the speed and low hardware demands of a lookup table
technique that is working on the entire shape space of droplets and not
just a few selected shape parameters.

The article is organized as follows. In Sec.\ \ref{sec:physics}
we first address the underlying physics of pendant drops and
present a derivation of
the shape equations that a pendant drop needs to fulfill.
We also classify all possible pendant drop shapes
under pressure and volume control to find the experimentally
relevant shapes and the parameter regimes where they exist in nature.
Numerically solving the forward problem for the relevant shapes
provides the basis for the design and training of a deep neural network
that solves the inverse problem to determine the surface
tension from a pending drop shape
as indicated in Fig.\ \ref{fig:network}. This machine learning approach
is presented in Sec.\ \ref{sec:ML}.  Results from conventional
shape fitting and machine learning tensiometry are compared in
Sec.\ \ref{sec:results}.

\section{Physics of pendant drops}
\label{sec:physics}

\subsection{Arc length parametrization}

\begin{figure}[h]
  \includegraphics{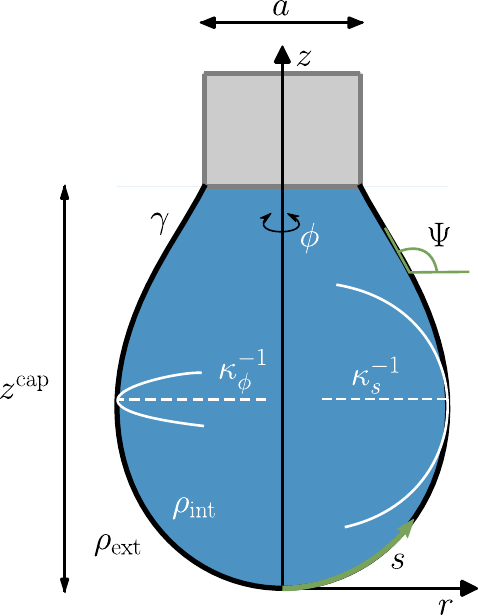}
  \caption{Visualization of a liquid drop in arc length parametrization.}
  \label{fig:parametrization}
\end{figure}

A sensible parametrization of an axisymmetric hanging droplet shape is the arc
length parametrization for which the first two shape equations can be found by
purely geometric arguments:
\begin{align}
  \label{eqn:first-shape-equation}\frac{\mathrm{d}r}{\mathrm{d}s}
  &= \cos \Psi \,,
    \\
  \label{eqn:second-shape-equation}\frac{\mathrm{d}z}{\mathrm{d}s}
  &= \sin \Psi \,,
\end{align}
where we use cylindrical coordinates $(r,z)$ with
the $z$-axis as the axis of symmetry and $\Psi$ is the angle
of the drop normal with the $z$-axis, see Fig.\ \ref{fig:parametrization}.
The principal curvatures in this parametrization are given by the
circumferential curvature $\kappa_\phi = {\sin \Psi}/{r}$ and the
meridional curvature $\kappa_s = {\mathrm{d}\Psi}/{\mathrm{d}s}$.

The
boundary conditions for a drop hanging from a capillary with diameter $a$ are
given by $r(s = 0) = 0$, $\Psi(s = 0) = 0$, $z(s = 0) = 0$ and
$r(s = L) = {a}/{2}$,
where $L$ is the total length of the arc. The boundary conditions at $s = 0$
describe the apex of the drop, where the radius, the arc angle and the height
are fixed to zero. Only the last boundary condition at $s = L$ describes the
attachment to the capillary.

\subsection{Young-Laplace equation from local force balance}

We consider a droplet with a
density difference $\Delta \rho = \rho_\mathrm{int}- \rho_\mathrm{ext}$
across the interface,  attached to the capillary and pulled down
by gravity.
The problem can be discussed for given Laplace  pressure
$p_\mathrm{L}$ at the apex of the drop
or prescribed drop volume $V$.
For prescribed drop volume,
$p_\mathrm{L}$ is introduced as a Lagrange multiplier to fulfill
the volume constraint.
In both cases the Young-Laplace equation follows from a vanishing
first variation of the droplet energy which is a necessary condition
that stationary droplet shapes have to fulfill.

There are several ways to derive the Young-Laplace equation based on the
concept of energy minimization or, equivalently, local force balance.
Here, we consider the forces along the $z$-axis.
We cut the drop at height $z$ and consider the
 $z$-components of the total forces
on the lower part of the drop.
There are four forces acting on every horizontal slice of the drop,
the surface tension force component in $z$-direction $F_\gamma^z$, the pressure
force $F_p$, the gravitational force $F_g$ caused by the mass hanging below
height $z$ and the buoyancy force caused by the difference in density $F_B$
\begin{align}
  F_\gamma^z(z) &= 2 \pi r(z) \gamma \sin \Psi \\
  F_p(z) &= - p(z)\pi r^2(z) \\
  F_g(z) + F_B(z) &= -\Delta m(z) g
\end{align}
with
\begin{equation}
  \Delta m(z) = \pi \Delta \rho \int_0^z \mathrm{d}z' r^2(z') \,,
\end{equation}
the  mass difference below height $z$
and the hydrostatic pressure $p(z)= p_\mathrm{L} - \Delta\rho g z$,
where $p_\mathrm{L}$ is  the pressure at the apex of the drop. The force
balance condition then states at any height $z \in [0, z^{\mathrm{cap}}]$:
\begin{equation}
 p(z)\pi r^2(z) =  2 \pi r(z) \gamma \sin \Psi - m(z)g\,.
  \label{eqn:force-balance}
\end{equation}
Taking the derivative ${\mathrm{d}}/{\mathrm{d}z}$
on both sides
\begin{equation*}
  \frac{\mathrm{d}}{\mathrm{d}z}\left( r^2(z) p(z) \right)
 = 2 \gamma \frac{\mathrm{d}}{\mathrm{d}z} \left( \kappa_\phi(z) r^2(z) \right) - g\rho r^2(z)
\end{equation*}
and using
\begin{align*}
  \frac{\mathrm{d} \kappa_\phi}{\mathrm{d}z}
  &= \frac{\cot\Psi(z)}{r(z)}\left(\kappa_s - \kappa_\phi \right) \\
  \frac{\mathrm{d} r}{\mathrm{d}z} &= \cot(\Psi)
\end{align*}
leads to the Young-Laplace equation
\begin{equation}
 p(z)=  p_\mathrm{L} - \Delta\rho g z = \gamma (\kappa_s + \kappa_\phi) \,,
  \label{eqn:young-laplace}
\end{equation}
where the interfacial tension $\gamma$, the apex pressure
$p_\mathrm{L}$,
and the density difference across the interface $\Delta\rho$ are
constant along the interface.
Vice versa, the force balance  $F_\gamma^z(z)+F_g(z) + F_B(z) +F_p(z)=0$ from
\eqref{eqn:force-balance}  is a first integral of the
Young-Laplace equation.
At the apex, we have $\kappa_s=\kappa_\phi$ by axisymmetry. Therefore,
the apex Laplace pressure $p_\mathrm{L}$ is experimentally observable via the
radius of curvature $R_0$ in the apex, $p_\mathrm{L}= 2\gamma/R_0$.

Inserting $\kappa_s$ and $\kappa_\phi$ as the principal curvatures into
\eqref{eqn:young-laplace} leads to the final shape equation of the pendant
drop:
\begin{equation}
  \frac{\mathrm{d}\Psi}{\mathrm{d}s} =
  \frac{p_\mathrm{L}}{\gamma}
   - \frac{\Delta\rho g z}{\gamma} - \frac{\sin \Psi}{r} \,.
  \label{eq:shape3}
\end{equation}
Shape equation \eqref{eq:shape3} has a numerical singularity at $r(s = 0) = 0$,
which can be circumvented by applying de L'H\^ospital's  rule and using the
axisymmetry in the apex, yielding the
limit $\mathrm{d} \Psi / \mathrm{d}s (s \to 0) \to  p_\mathrm{L} / 2 \gamma$.

Solutions to the shape equations with $z(0)=0$ and the attachment
boundary condition $r(s = L) = {a}/{2}$ will have a variable
droplet height $z_\mathrm{cap} = z(s=L)$ and, thus, also a variable
pressure $p_\mathrm{cap} = p_\mathrm{L} - \Delta\rho g z_\mathrm{cap}$
at the capillary. While  the  apex pressure $p_\mathrm{L}$
is experimentally observable via the apex curvature and
a theoretically convenient control parameter, the experimental
situation is usually such that the capillary
is at a fixed position (i.e., $z_\mathrm{cap}$ is fixed)
and, if working under pressure control,
the capillary pressure $p_\mathrm{cap}$ is controlled rather than
the apex pressure $p_\mathrm{L}$.

\subsection{Non-dimensionalization and control parameters}

We choose the  length scale $a$ for non-dimensionalization
as the diameter of the capillary
leading to the definitions $\tilde{z} \equiv {z}/{a}$,
$\tilde{r} \equiv {r}/{a}$, $\tilde{s} \equiv {s}/{a}$, and
$\tilde{\kappa}_{s, \phi} \equiv {\kappa_{s, \phi}}{a}$.
The non-dimensional
form of the Young-Laplace equation
\eqref{eqn:young-laplace} is  given by
\begin{align}
  \tilde{p}_\mathrm{L} - \Delta\tilde{\rho}\tilde{z} &=
  \tilde{\kappa}_s + \tilde{\kappa}_\phi ~~\mbox{with}
  \\
  \tilde{p}_\mathrm{L} &\equiv \frac{p_\mathrm{L} a}{\gamma}  ~~\mbox{and}~~
      \Delta\tilde{\rho} \equiv    \frac{\Delta \rho g a^2}{\gamma},
  \label{eqn:nondimensional_laplace}
\end{align}
where we
introduced the non-dimensional apex pressure
$\tilde{p}_\mathrm{L}$ and the non-dimensional
gravitational control parameter $\Delta\tilde{\rho}$.

Note that setting the length scale for non-dimensionalization
to the radius of curvature in the apex
of the drop $R_0$ further eliminates the non-dimensional apex pressure from the
system of differential equations,
since ${p_\mathrm{L} R_0}/{\gamma} = 2$,
leading to the often used definition of the bond number \cite{Berry2015,
  Saad2011, Stauffer1965, Morita2002}
\begin{equation}
  \mathrm{Bo} = \frac{\Delta \rho g R_0^2}{\gamma} = \frac{4\Delta\rho g
    \gamma}{p_\mathrm{L}^2}
\end{equation}
as a single non-dimensional control parameter.  For free-standing droplets
without attachment to a capillary the bond number $\mathrm{Bo}$ is the only
shape control parameter.  As soon as an attachment boundary condition, e.g.,
$r(s= L) = {a}/{2}$,
is applied  a second
control parameter must be defined, which involves
the attachment length scale $a$.
When using $R_0$ for non-dimensionalization  this additional
control parameter is hidden in the attachment boundary condition
itself. We choose the non-dimensionalization
length scale $a$ such  that
the attachment boundary condition is parameter-free and, thus, get
the Laplace pressure
$\tilde{p}_\mathrm{L}$ in the apex
and the dimensionless density difference
$\Delta \tilde{\rho}={\Delta \rho g a^2}/{\gamma}$,
which can also be interpreted as a dimensionless
measure for the square of the capillary diameter, as two independent
non-dimensional shape control parameters.
For water droplets in air with $\gamma = 72 {\rm mN/m}$,
a value $\Delta \tilde{\rho}=1$ corresponds to a
capillary diameter of $a= 2.7 {\rm mm}$.

Note that we limit our focus to pendant drops, so $\Delta \tilde{\rho}$
is always positive. When considering setups where the drop rises from a
capillary $\Delta \tilde{\rho}$ can also be negative.

From fitting the pendant drop
shape (either conventionally or by machine learning)
we will obtain a guess for the two dimensionless parameters
$\tilde{p}_\mathrm{L}$ and $\Delta \tilde{\rho}$. If pressure
is not measured in the experiment, the surface tension has to be
extracted from the parameter  $\Delta \tilde{\rho}$ for known
density contrast $\Delta \rho$ and capillary diameter
$a$.
In this sense, $\Delta \tilde{\rho}$ is the more important parameter
to determine. From the second parameter $\tilde{p}_\mathrm{L}$, we can
then obtain a measurement of the actual apex pressure.

\subsection{Droplet shapes  classified by bulges and necks}

\begin{figure}
   \includegraphics[width=0.99\linewidth]{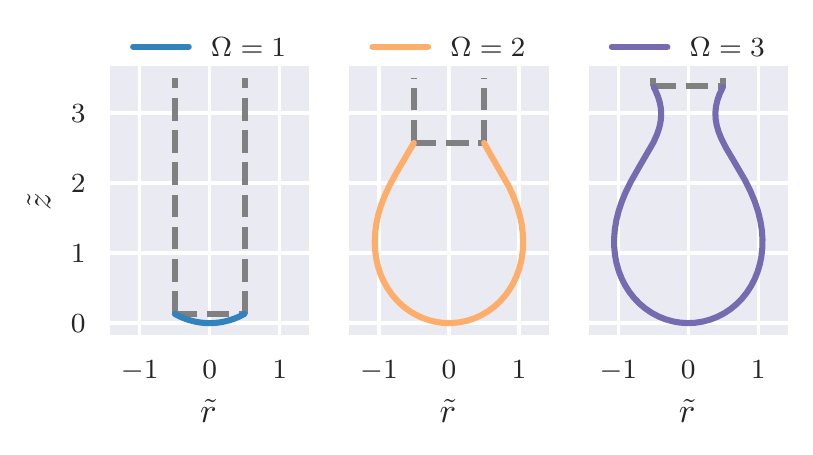}
  \caption{
    Comparison of valid solution shapes for solution classes
    $\Omega \in \{1, 2, 3\}$ for parameters: $\tilde{p}_\mathrm{L} = 2$,
    $\Delta \tilde{\rho} = 0.3$. The dashed grey lines indicate the
    capillary. Higher class solutions always contain the shapes of all
    lower class solutions, since all classes are constructed
    from the same general solution
    shape that is cut off at different heights.
  }
  \label{fig:solution_classes}
\end{figure}

We will discuss droplet shapes either under apex pressure control
(parameter $\tilde{p}_\mathrm{L}$) or under volume control (with
$p_\mathrm{L}$ as a Lagrange multiplier).
Even these two non-dimensional shape control parameters are not sufficient
to fully
characterize the pendant drop's shape.
The Young-Laplace equation with height-dependent hydrostatic
pressure \eqref{eqn:young-laplace} has no closed analytical
solutions; solutions for pendant drops are
distorted unduloids \cite{Padday1971}.
An unduloid is an axially symmetric
constant mean curvature surface with
a curvature ratio
$|\kappa_s/\kappa_\phi|< 1$. This curvature condition is also fulfilled for
the droplet profiles, but the mean curvature is decreasing for $z>0$
because of the decreasing hydrostatic pressure $p(z)$.
Therefore, similarly to an unduloid,  the droplet profile
 radial distance function   $\tilde{r}(\tilde{s})$
contains several maxima (bulges) and minima (necks) for larger
$\tilde{p}_\mathrm{L}$,
such that   the attachment
boundary condition may be fulfilled at a number of different total
dimensionless arc lengths $\tilde{L}$ along the same  solution of the
shape equations leading to different shapes for the same
choices of $\tilde{p}_\mathrm{L}$ and $\Delta \tilde{\rho}$.
This gives rise to  several possible classes of
solution shapes which can be characterized by their number
of bulges and necks and the first three of which are shown
in Fig.\ \ref{fig:solution_classes}.
The number of bulges and necks
is counted by another  discrete parameter
\begin{equation}
  \Omega \equiv 1+ \# \mathrm{necks} +  \# \mathrm{bulges}
  \label{eq:Omega}
\end{equation}
that indicates
the class of a solution.

The first class of solutions, $\Omega = 1$, is a simple convex shape with
$ \tilde{r}(\tilde{s}) < {1}/{2}$ for all $ 0\le  \tilde{s} < \tilde{L}$;
this class has a  monotonically increasing radius with
$\tilde{r}(\tilde{s} = 0) = 0$ in the apex and
$\tilde{r}(\tilde{L}) = {1}/{2}$ at the capillary.

The second
class of solutions, $\Omega = 2$, are convex shapes for which
there exists exactly one  bulge, where we define  a bulge  as a
point  where
$\tilde{r}(\tilde{s})$ has a local
maximum (such that  $\sin\Psi(\tilde{s})=1$
see \eqref{eqn:first-shape-equation}).
The  $\Omega = 2$ shapes are convex and always bulge out, i.e.,
the bulge is  wider than the capillary.
The shape class $\Omega = 2$ will be the most important class
for shapes under volume control.

The third solution class,
$\Omega = 3$, is the first class of non-convex solutions.
These solutions have
exactly one bulge and one neck, where a neck is defined as a  point where
$\tilde{r}(\tilde{s})$ has a local
minimum (such that also  $\sin\Psi(\tilde{s})=1$,
see \eqref{eqn:first-shape-equation}). The $\Omega = 3$ solutions
have a neck at the capillary and always
cross the capillary boundary condition from left
to right, i.e.,
${\mathrm{d}\tilde{r}(\tilde{s})}/{\mathrm{d}\tilde{s}} \rvert_{\tilde{s}
  = \tilde{L}} \ge 0$.

Continuing this scheme there also exist higher classes $\Omega>3$ of
shapes, in principle, which are characterized by their increasing number
of bulges and necks.
While all shape classes up to $\Omega=3$ can  actually be
observed in experiments,  higher classes $\Omega>3$
are not  observed because they are energetically unfavorable
and unstable both under volume and pressure control as we will show
in the next section.

\subsection{Shape bifurcations and shape diagram for pendant drops}

\begin{figure*}
  \includegraphics[width=0.99\linewidth]{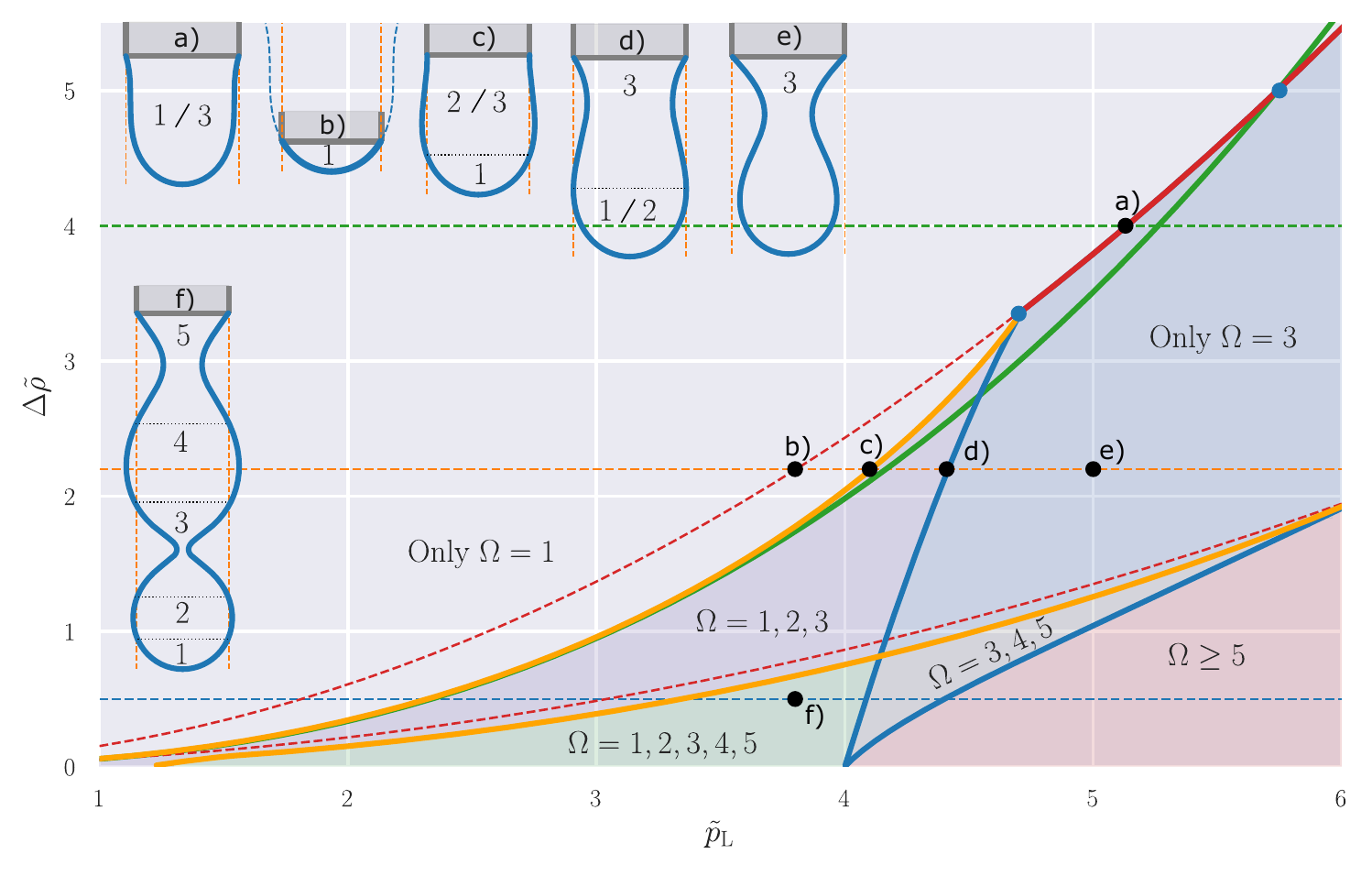}
  \caption{
    Shape diagram in the
    $\tilde{p}_\mathrm{L}$-$\Delta \tilde{\rho}$ parameter plane
     for apex pressure control.
    Shapes corresponding to black points in parameter space are
    shown on the right.
    At the yellow lines two additional droplet shapes $\Omega =n+1,n+2$
    appear,
    at the blue lines two droplet shapes $\Omega = n,n+1$
    annihilate in bifurcations ($n=1,3,...$).
    Blue and yellow lines terminate in a critical point at
    $\Delta \tilde{\rho}=3.37$ for $n=1$ and
    $\Delta \tilde{\rho}=2.07$ for $n=3$.
    At the red lines shapes develop saddle points; the lower pointed
    part of the red line is not observable as the saddle develops
    in the continuation of the shape to the region above
     the capillary opening and outside the capillary. At the
    green line, the maximal volume is reached for increasing pressure.
     Droplets detach  at this line,
    either in shape $\Omega=1$ for $\Delta \tilde{\rho}>5.02$
    or in shape $\Omega=3$ for $\Delta \tilde{\rho}<5.02$.
    Example shapes a)-e) illustrating the bifurcations
     are shown for $\Delta \tilde{\rho}=4$
    (green dashed line, shape a), $\Delta \tilde{\rho}=2.2$
     (yellow dashed line, shapes b-e), and  $\Delta \tilde{\rho}=0.5$
     (blue dashed line, shape f).
  }
  \label{fig:shapediagram}
\end{figure*}

\begin{figure}
  \includegraphics[width=0.99\linewidth]{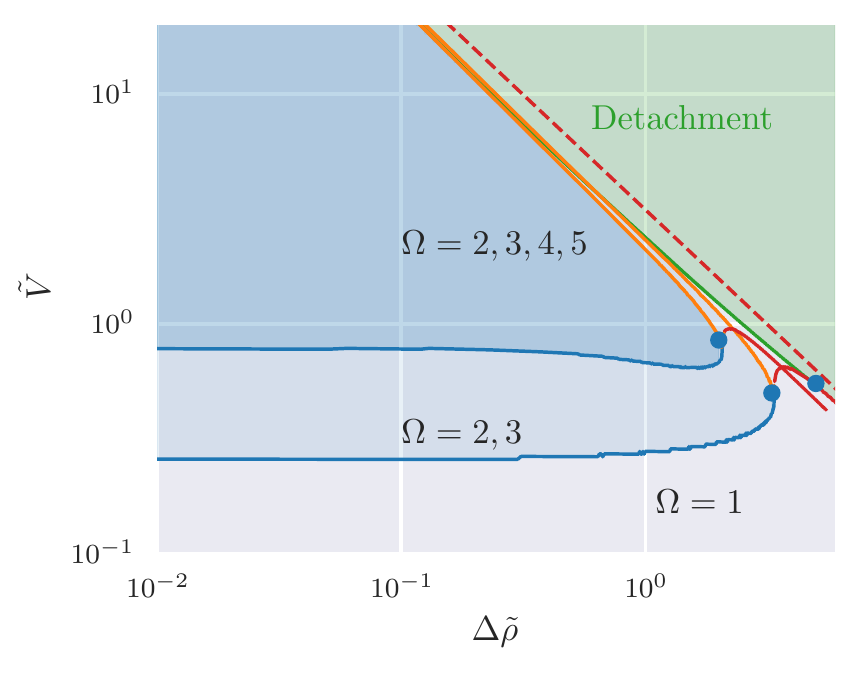}
  \caption{
    Shape diagram in the
    $\Delta \tilde{\rho}$-$\tilde{V}$  parameter plane for
    volume control.
    At small volumes  only shape $\Omega=1$ is possible.
    Almost up to the detachment volume both shapes
    $\Omega=2,3$ are possible. Where both shapes (or even higher classes)
    are possible,
    the bulged shape $\Omega=2$ is the global energy minimum.
     Only in a small region
     close to the maximal volume, the necked
     shape $\Omega=3$  is the global energy minimum.
     The upper green line marks the
     maximal volume, where the droplet detaches.
     For $\Delta \tilde{\rho}>5.02$ it detaches in shape $\Omega=1$,
      for $\Delta \tilde{\rho}<5.02$  in
      the necked shape  $\Omega=3$.
    According to the approximative Tate's law detachment
    happens at $\tilde{V} = \pi/\Delta \tilde{\rho}$ (dashed red line);
  }
  \label{fig:shapediagramV}
\end{figure}

For the tensiometry analysis, we first discuss
where the different shape classes can be found in the
$\tilde{p}_\mathrm{L}$-$\Delta \tilde{\rho}$ parameter plane
under pressure control and in the
$\Delta \tilde{\rho}$-$\tilde{V}$ parameter plane under volume control,
which leads to the shape diagrams
Figs.\ \ref{fig:shapediagram} and \ref{fig:shapediagramV}.
This will be important for identifying the experimentally
relevant parameter regions and  to rationalize parameter sensitivity
of shapes and the
selection of the relevant shapes for the training of the neural
network.

We will first discuss all possible
shapes under apex pressure control.
This means we
integrate the shape equations (\ref{eqn:first-shape-equation}),
(\ref{eqn:second-shape-equation}), and (\ref{eq:shape3}) in dimensionless
form starting at the apex with  given $\tilde{p}_\mathrm{L}$ and
$\Delta \tilde{\rho}$ and ignoring the attachment
boundary condition at the capillary.
At every intersection with the capillary,
where $\tilde{r}(\tilde{s}) = 1/2$, the remaining attachment
boundary condition can be fulfilled with a different
arc length. This means for a solution which intersects
$n$-times with the capillary, all shape classes
$\Omega = 1,...,n$ can occur in the shape diagram for this choice
of parameters.

For small pressure, there is only one intersection and
only shapes with $\Omega=1$ exist.
For increasing apex pressure $p_\mathrm{L}$ the curvature
of droplet shapes increases and higher order shapes $\Omega>1$
with more bulges and necks
become possible in a sequence of bifurcations, which
are shown in the bifurcation diagram
Fig.\ \ref{fig:shapediagram}.

We follow the sequence of bifurcations for fixed $\Delta \tilde{\rho}$
and increasing apex pressure $p_\mathrm{L}$.
In a first simple fold bifurcation
a bulge and neck pair is
formed  via a saddle point configuration of the droplet.
At this bifurcation, an  $\Omega=n$ shape transforms
into an  $\Omega=n+2$ shape.
The first bulge/neck pair is
formed at  the left red bifurcation  line in  the shape diagram
Fig.\ \ref{fig:shapediagram} (transition $\Omega=1\to 3$);
higher order lines exist but are not shown.
We find numerically that the red dotted bifurcation lines are parabolas with
$\Delta \tilde{\rho} \simeq 0.15\,  \tilde{p}_\mathrm{L}^2$
and $\Delta \tilde{\rho} \simeq 0.054\,  \tilde{p}_\mathrm{L}^2$.

There is a  critical value
$\Delta \tilde{\rho}_\mathrm{c1} \simeq  3.37$, where
the first saddle forms
exactly at the capillary radius $\tilde{r}=1/2$.
For
$\Delta \tilde{\rho}> 3.37$ the first saddle forms at a radius smaller than the
capillary radius (at $\tilde{r}<1/2$,
see Fig.\ \ref{fig:shapediagram}, shape a), for $ \Delta \tilde{\rho}< 3.37$
it would form above the capillary opening and ``outside'' the capillary
(see Fig.\ \ref{fig:shapediagram}, shape b), i.e., the saddle would
form in the region $\tilde{r}(s)>1/2$ and for $s>L$
above the only possible attachment
point to the capillary, where $\tilde{r}(L)=1/2$).
Therefore, the bifurcation at the
red line is unobservable in an actual experiment
for  $ \Delta \tilde{\rho}< 3.37$ (dotted red line).
Likewise,
there is a critical value $ \Delta \tilde{\rho}_\mathrm{c2} \simeq  2.07$
for the formation of the second saddle, and for
$ \Delta \tilde{\rho}> 2.07$ the second  saddle forms inside the
capillary, while it forms above and outside for $ \Delta \tilde{\rho}< 2.07$
(dotted red line).
These critical values also exist for the higher order lines, in principle.
The critical values $ \Delta \tilde{\rho}_\mathrm{c1} \simeq  3.37$
and $ \Delta \tilde{\rho}_\mathrm{c2} \simeq  2.07$
define critical points
on the  respective bifurcation lines.
Because a saddle configuration has vanishing curvature $\kappa_s=0$,
a saddle configuration right at the capillary
has a capillary Laplace pressure
$\tilde{p}_\mathrm{cap} =  \tilde{\kappa}_\phi = 2$, which
is thus the capillary pressure for all critical points.

For capillary widths smaller than
the critical values   ($ \Delta \tilde{\rho}< 3.37$ for the
first  bulge/neck pair or
$ \Delta \tilde{\rho}< 2.07$ for the second),
bifurcations occur only
after a bulge/neck pair has formed ``outside'' the capillary
(at  $\tilde{r}>1/2$). Then  bulge and neck
``move inwards'' towards the symmetry axis upon increasing the pressure
further, and
droplet shapes bifurcate
if a neck moves inwards and touches the capillary radius (i.e., the
neck is at $\tilde{r}=1/2$, see shape c in Fig.\ \ref{fig:shapediagram}).
Coming into this bifurcation with the highest
possible shape $\Omega=n$ ($n=1,3,...$ odd),
a pair  $\Omega= n+1,n+2$  of  additional  shapes  become
possible  in a  simple fold  bifurcation (at the
yellow bifurcation lines in  the shape diagram
Fig.\ \ref{fig:shapediagram}, illustrated for $n=1$ with shape c).
Right at these  bifurcation  lines,
both shapes $\Omega= n+1,n+2$
are identical and
the droplet has a vertical tangent at the capillary.

Likewise,
if a bulge moves inwards and
touches the capillary, a pair  $\Omega= n,n+1$  of shapes
annihilates again in a simple fold bifurcation (at the
blue bifurcation lines in  the shape diagram
Fig.\ \ref{fig:shapediagram}, illustrated for $n=1$ with shape d).
Beyond the blue bifurcation lines
$\Omega=n+2$  is the lowest possible order of shapes.
Right at these bifurcation lines,
both shapes $\Omega= n,n+1$
are identical and also have
a vertical tangent at the capillary.
As a result,
in the magenta and green shaded areas between
the yellow and blue bifurcation lines, classes $\Omega=1,2,3$
are possible, in the green and grey shaded areas classes  $\Omega=3,4,5$ are
possible, and so on (see shape f in Fig.\ \ref{fig:shapediagram}).

At the first blue bifurcation line, where shapes
$\Omega= 1,2$ annihilate, these shapes
are approximately half-spherical (see shape d in  Fig.\ \ref{fig:shapediagram}).
They are exactly
half-spherical  for $\Delta \tilde{\rho}=0$
with radius  $\tilde{R}_0 = 2/\tilde{p}_\mathrm{L}=1/2$, and
volume $\tilde{V} = \pi/12$.
Along the blue bifurcation line, the maximal
dimensionless Laplace pressure increases to $\tilde{p}_\mathrm{L}>4$
for $\Delta \tilde{\rho}>0$,
 because the
shape elongates and the apex acquires a  higher curvature.

Both at the yellow and  blue  bifurcation lines
(for example for shapes c and d in  Fig.\ \ref{fig:shapediagram}),
the force equilibrium
(\ref{eqn:force-balance}) holds at $z=z_\mathrm{cap}$
with $r(z_\mathrm{cap})=a/2$
and   $\sin\psi=1$ (vertical tangent) resulting in the
exact bifurcation condition
\begin{equation}
  \Delta \tilde{\rho}\frac{\tilde{V}}{\pi} =
   1 - \frac{\tilde{p}^\mathrm{cap}}{4}
  \label{eqn:bifurcation_cond}
\end{equation}
which holds along the entire boundary of
the magenta, green and grey shaded areas
in the shape diagram Fig.\ \ref{fig:shapediagram}.

The birth of  bulge/neck pairs outside the capillary radius
(at the dotted red lines), and their
subsequent inward motion with increasing pressure
with first the neck crossing the capillary radius
(at the yellow bifurcation lines)
and then the bulge moving through the capillary radius (at the blue
bifurcation lines)
explains the structure of the
pressure shape diagram Fig.\ \ref{fig:shapediagram}
for  $ \Delta \tilde{\rho}< 3.37$, i.e., for sufficiently
narrow capillaries.
Here we have several possible
shape sequences $\Omega = 1\to 1,2,3 \to ...$ (see shapes b to e or shape f
in  Fig.\ \ref{fig:shapediagram}).
For $\Delta \tilde{\rho}> 3.37$ or wide capillaries,
the birth of the first
bulge/neck pair inside the capillary radius (at the solid red line,
see shape a in  Fig.\ \ref{fig:shapediagram})
with $\Omega=1 \to 3$
is the only bifurcation event.

We can obtain a corresponding volume shape diagram   in the
$\Delta \tilde{\rho}$-$\tilde{V}$  parameter plane, which is
shown in Fig.\
\ref{fig:shapediagramV}. Again, for  $\Delta \tilde{\rho}< 3.37$
there are several possible shape sequences $\Omega= 1\to 2,3 \to ...$,
whereas for  $\Delta \tilde{\rho}> 3.37$, there is only one bifurcation
 $\Omega=1 \to 3$ possible.

In the shaded areas in the shape diagrams between
the yellow and blue bifurcation lines, several shape classes $\Omega$
are possible. Which shape class is actually assumed
because it is stable and energetically  favorable and
which class is
only metastable, depends on the dimensionless
energy (measured in units of $a \gamma$)
\begin{align}
  \tilde{F} &=  \int_0^{\tilde{L}}
  \mathrm{d} \tilde{s} \left(2\pi  \tilde{r}(\tilde{s})
  + \pi \Delta \tilde{\rho}  \tilde{z}(\tilde{s})
   \tilde{z}'(\tilde{s})\tilde{r}^2(\tilde{s})\right)
\end{align}
of the shape
for volume control or the enthalpy
\begin{align}
  \tilde{G} &= \tilde{F} - \tilde{p}_\mathrm{L} \tilde{V}
     = \tilde{F} - \tilde{p}_\mathrm{L}  \pi \int_0^{\tilde{L}}
  \mathrm{d} \tilde{s} \tilde{z}'(\tilde{s})
  \tilde{r}^2(\tilde{s})
\end{align}
for apex pressure control.
The first term in $\tilde{F}$ is the surface energy, the second
term the gravitational energy (measured with respect to the
apex $\tilde{z}(0)=0$).
At fixed volume,  $\tilde{F}$ is minimized in a stationary state,
while at fixed apex pressure $\tilde{p}_\mathrm{L}$, the enthalpy
$\tilde{G}$ is extremized resulting in $\tilde{p}_\mathrm{L} =
\mathrm{d}\tilde{F}/\mathrm{d} \tilde{V}$.
If several shape classes are possible, the
shape with the minimal energy  $\tilde{F}$ is stable for
volume control, while the shape with minimal enthalpy
$\tilde{G}$ is stable for apex pressure control.

A  shape must have
$\mathrm{d}\tilde{p}_\mathrm{L}/\mathrm{d} \tilde{V}>0$ to be stable
under apex pressure control, otherwise it could increase volume
without limit at a given maintained pressure.
This implies  a convex energy  $\tilde{F}(\tilde{V})$
or a concave enthalpy $\tilde{G}(\tilde{p}_\mathrm{L})$
as necessary stability condition under pressure control.
Stability under volume  control can be deduced from
the properties of the $\tilde{p}_\mathrm{L}(\tilde{V})$-relation
using the criteria derived by Maddocks \cite{Maddocks1987}.
Thus it is necessary to know the  $\tilde{G}(\tilde{p}_\mathrm{L})$-,
$\tilde{F}(\tilde{V})$-,  and  $\tilde{p}_\mathrm{L}(\tilde{V})$-relations
to decide on the stability of shapes.

\begin{figure*}
  \includegraphics[width=0.99\linewidth]{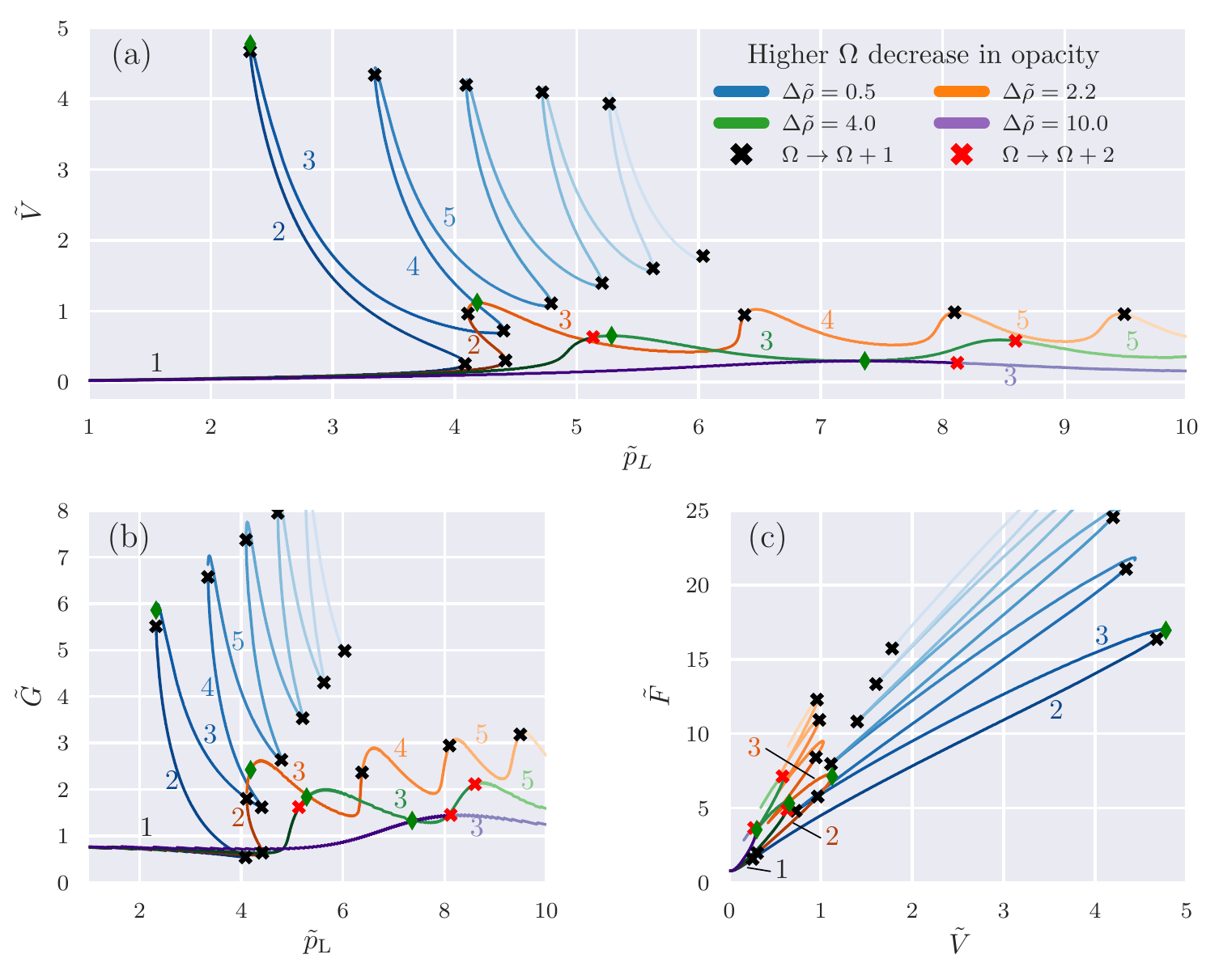}
  \caption{
    (A) Volume-pressure relationship,  (B) enthalpy as a function of pressure,
    and (C)  energy as a function of volume
    for
    four shapes with  $\Delta \tilde{\rho}=0.5$ (blue)
    $\Delta \tilde{\rho}=2.2$ (yellow)
    $\Delta \tilde{\rho}=4.0$ (green), and  $\Delta \tilde{\rho}=10.0$
    (red).
    We marked where the bifurcations $\Omega=1\to 1,2,3$ (yellow line
     in Fig.\ \ref{fig:shapediagram}) and
    $\Omega=1,2,3\to 3$  (blue line
     in Fig.\ \ref{fig:shapediagram}) occur, and the shape of maximal volume.
  }
  \label{fig:pV}
\end{figure*}

Therefore,
we follow the evolution of all shapes in
terms of  apex pressure $\tilde{p}_\mathrm{L}$, droplet volume $\tilde{V}$
    and droplet energy $\tilde{F}$ as well as enthalpy $\tilde{G} = \tilde{F}
    -\tilde{p}_\mathrm{L} \tilde{V}$
    through all bifurcations
    in Fig.\ \ref{fig:pV}
    for the four values $\Delta \tilde{\rho}=0.5,~2.2,~4.0,~10.0$
(also indicated in the shape diagram Fig.\ \ref{fig:shapediagram})).
The bifurcation points, where  shapes $\Omega =1,2$ vanish and where
additional shapes $\Omega =2,3$ appear are marked with black dots,
the shape with  maximal volume with a green dot.

For $\Delta \tilde{\rho}=4.0,~ 10.0$, i.e.,
wide capillaries $\Delta \tilde{\rho} > 3.37$ there is only
a single bifurcation $\Omega=1\to 3$ of the shape upon increasing
the pressure, namely, when a bulge/neck pair is
created  (marked with a red dot in Fig.\ \ref{fig:pV} corresponding to the
red line in the shape diagram Fig.\ \ref{fig:shapediagram}).
Fig.\ \ref{fig:pV} shows that the  $\tilde{p}_\mathrm{L}(\tilde{V})$-curve
is monotonously increasing up to the maximal volume.
Therefore shapes $\Omega=1$ and  $\Omega=3$ are stable
under pressure and volume control up to the shape of maximal volume.
The maximal volume shape  is attained in a shape $\Omega=3$ for
$\Delta \tilde{\rho}=4.0$ but in shape $\Omega=1$ for
$\Delta \tilde{\rho}=10.0$.

For $\Delta \tilde{\rho}=2.2$, i.e., in the regime
$2.07 < \Delta \tilde{\rho}< 3.37$ of narrower capillaries,
there are  two observable
bifurcations
upon increasing the pressure.
Shapes $\Omega =2,3$ appear
when a neck appears at the capillary
(yellow bifurcation line in  Fig.\ \ref{fig:shapediagram}),
and shapes  $\Omega =1,2$ meet and vanish  when  a bulge appears
at the capillary (blue bifurcation
line in  Fig.\ \ref{fig:shapediagram}).
Both bifurcations are marked  by black dots in Fig.\ \ref{fig:pV}.
 The  $\tilde{p}_\mathrm{L}(\tilde{V})$-curve
 is monotonously increasing for shape $\Omega=1$  up to the
 bifurcation, where  shapes $\Omega =1,2$ vanish.
 In addition, there is an increasing  $\tilde{p}_\mathrm{L}(\tilde{V})$-curve
 for shape  $\Omega=3$  up to the shape with maximal volume (green dot).

For small  $\Delta \tilde{\rho}<2.07$ even five or more
shapes can coexist in certain parameter regimes.
For  $\Delta \tilde{\rho}=0.5$, there is a sequence
$\Omega=1 \to 3 \to 1,2,3 \to 3,4,5 \to 1,2,3,4,5 \to 3,4,5 \to 5$
via  six bifurcations.
 The  $\tilde{p}_\mathrm{L}(\tilde{V})$-curve
 is monotonously increasing for shape $\Omega=1$  up to the
 bifurcation, where  shapes $\Omega =1,2$ vanish. For all
 higher shapes the  $\tilde{p}_\mathrm{L}(\tilde{V})$-curves are
 almost everywhere decreasing except for very small
 pieces around the maximal volumes of these shapes.

 We conclude that $\Omega=1$ is the only shape which always has an
 increasing $\tilde{p}_\mathrm{L}(\tilde{V})$-relation and
 is generally stable under pressure and volume control.
 Under pressure control, shape  $\Omega=1$  always has the lowest
 enthalpy $G$ and is the energetically preferred state where it exists
 (see Fig.\  \ref{fig:pV}(B)).
 Because the  $\tilde{p}_\mathrm{L}(\tilde{V})$-curves are
 S-shaped, we can deduce from the theorems derived by Maddocks
 \cite{Maddocks1987} that shape $\Omega=2$ is unstable
 under pressure control but stable under volume control.
 Under volume control, shape  $\Omega=2$  always has the lowest
 energy  and is the energetically preferred state where it exists
 (see Fig.\  \ref{fig:pV}(C)).
 Shape $\Omega=3$ is stable under pressure and volume control
 in a small regime from the bifurcation $1\to 1,2,3$ where it appears
 together with shape 2
 up to the maximal volume (green dots in Fig.\ \ref{fig:pV});
 in this regime it has an increasing
 $\tilde{p}_\mathrm{L}(\tilde{V})$-relation and is the energetically
 preferred state under volume control.
 Beyond the shape of maximal volume,
 shape $\Omega=3$  becomes unstable under pressure
 control but remains metastable under volume control.
 Shapes $\Omega\ge 3$
 are, however, always
 energetically unfavorable as higher order
 shapes have higher energy and enthalpy,
 as can be seen in Figs.\ \ref{fig:pV}(B,C).
 Under volume control, shapes $\Omega=1,2,3$ are stable
 up to the maximal volume. All higher order shapes $\Omega\ge 3$
 beyond the maximal volume are energetically unfavorable as
 Figs.\ \ref{fig:pV}(B,C) show.

 Experimentally, the standard situation is volume control.
 For this situation the sequence  $\Omega=1\to 2\to 3$  is the
 sequence of energetically
 preferred states with shape $\Omega=2$ being the global
 energy minimum in a large volume range (everywhere, where it exists).
 Therefore, we will focus on the shape classes $\Omega=2,3$
 in the tensiometry part of the paper.

\subsection{Droplets detach at the maximal droplet volume}

The pressure and volume shape diagrams are
limited by the maximally possible droplet volume
before detachment.
From Figs.\ \ref{fig:pV}(A,C) it is apparent that,
regardless of how complicated the bifurcation
sequence might be,
there always exists a maximal volume
$\tilde{V}_\mathrm{max}$ that a pendant drop can accommodate
for all values of $\Delta \tilde{\rho}$ (green dots in Fig.\ \ref{fig:pV}).
This maximal volume marks the end of existence of energetically
stable droplet shapes in the diagrams  Figs.\ \ref{fig:pV}(A,C).
This maximal volume is also essential for the stability under gravity.
If the droplet is loaded with more than the maximal volume, no
stationary state can exist, and the droplet has to start
moving downwards by gravity.
This leads to gravitational detachment of the droplet,
during which it dynamically breaks up into a stable pendant drop
of lower volume and a satellite droplet \cite{Eggers1994,Yildirim2005},
which is the basis of the drop weight method.
We only consider stationary droplet shapes and, thus, have only
access to the maximal stationary droplet volume {\it before} detachment
as it has also been used for tensiometry in Ref.\ \citenum{Gunde2001}.

We calculated the shapes of maximal volume numerically and
marked them by a green line in the shape diagram
in the $\tilde{p}_\mathrm{L}$-$\Delta \tilde{\rho}$ plane
in Fig.\ \ref{fig:shapediagram}.
This green line intersects the red line where the bifurcation
$\Omega=1\to 3$ occurs via formation of a bulge/neck pair.
This intersection happens at a value $\Delta  \tilde{\rho}\simeq 5.02$.
Therefore, the maximal volume
is attained in a shape $\Omega=3$ for  narrow capillaries
$\Delta \tilde{\rho}< 5.02$
and in a shape $\Omega=1$ for  wider capillaries
$\Delta \tilde{\rho}> 5.02$.

For small $\Delta \tilde{\rho} \lesssim 1$ the detachment (green line)
happens
almost at the same volume as the bifurcation where shapes $\Omega=2,3$
appear (yellow line) as can be seen both in  Figs.\ \ref{fig:shapediagram}
and \ref{fig:pV}. In this regime, detachment
happens with an almost vertical tangent at the capillary.
Therefore, the bifurcation condition (\ref{eqn:bifurcation_cond})
also gives an excellent description of the detachment
volume in this regime.
The similarity to the well-known Tate law is obvious: if we can approximate
$\tilde{p}^\mathrm{cap}\ll 4$, we recover  Tate's law \cite{Tate1864}
\begin{equation}
  \tilde{V} \approx \frac{\pi}{\Delta \tilde{\rho}}
  \label{eq:Tate}
\end{equation}
for gravitational detachment.
In the shape diagram Fig.\
\ref{fig:shapediagramV} in the
$\Delta \tilde{\rho}$-$\tilde{V}$  parameter plane,
Tate's law is shown as dashed
red line, the exact numerical detachment condition is
the green line. We clearly see that Tate's law overestimates the detachment
volume leading to the known underestimation of the surface tension by the
drop weight method \cite{Harkins1919, Garandet1994}.
We also observe in diagram \ref{fig:shapediagramV} that,
for  narrow capillaries
$\Delta \tilde{\rho}< 5.02$, droplets detach in a necked shape  $\Omega=3$.
while they detach in a simple  shape $\Omega=1$ for  wider capillaries
$\Delta \tilde{\rho}> 5.02$.

\section{Numerical approach}

In the following, we study two tensiometry approaches to extract
the two control parameters  $\tilde{p}_\mathrm{L}$ and $\Delta \tilde{\rho}$
from an experimental image of a droplet shape, conventional shape
fitting (CSF) as compared to a novel machine learning (ML) approach, where we
train a deep neural network to determine the control parameters.
To test both methods, we numerically generate droplet
shapes with known parameters  $\tilde{p}_\mathrm{L}$ and $\Delta \tilde{\rho}$
(the ``forward problem'') and then re-determine these
parameters by shape fitting or by the neural network (the ``inverse
problem'').
From now on we only consider shapes of the
classes $2$ and $3$, since they are predominantly used in tensiometry and
provide high fitting accuracy in general \cite{Berry2015}.
We start by only considering class $2$ shapes and discuss the generalized
approach to class $2$ and $3$ shapes afterwards.
This means we consider shapes in the  diagram
Fig.\ \ref{fig:shapediagram}, which lie in the magenta and green ``triangles''
enclosed by the yellow and blue bifurcation lines, where
shape classes $2$ can exist.

To numerically generate shapes for given
parameters $\tilde{p}_\mathrm{L}$ and $\Delta \tilde{\rho}$
we make use of a discretization of the shape
equations (\ref{eqn:first-shape-equation}),
(\ref{eqn:second-shape-equation}), and (\ref{eq:shape3})
to solve them iteratively in space.  For this a fourth order
Runge-Kutta
algorithm  is used, because it provides a good mix of accuracy and
speed. We use a modified version of \mbox{\emph{OpenCapsule}
  \cite{openCapsule, Hegemann2018}} for the numerical fitting and forward
solution of the Young-Laplace problem. The output data from the numerical
forward solution is evenly spaced in the arc lengths $\tilde{s}$ of the shape.

\section{Solving the inverse problem by conventional shape fitting}

The goal of shape fitting
is to numerically generate a shape that has the least square distance
to a set of sample points along the contour of an input shape.
The numerically generated optimal fit
will then make the parameters $\tilde{p}_\mathrm{L}$ and
$\Delta \tilde{\rho}$ of the input shape available.
In CSF, we start
with an initial guess for the parameters of the shape
$\{\tilde{p}_L^\text{initial}, \Delta
\tilde{\rho}_L^\text{initial}\}$. Second, we determine the Jacobian matrix for
the supplied parameters by giving every parameter a notch to either side,
comparing the resulting errors and numerically
calculating the derivatives this way. Last,
the parameters get updated with an update vector that points along the
steepest descent in error-parameter space. Generally, this is the way most
existing numerical implementations  perform the fitting.  After some
iterations a shape emerges that best fits the points from the contour of the
input shape for  a pair of best fitting material parameters
$\tilde{p}_\mathrm{L}$ and $\Delta \tilde{\rho}$.

\section{Machine learning approach for the inverse problem}
\label{sec:ML}

A ML approach provides a way to solve the computationally taxing
task of numerically fitting the shape in a more
efficient way by training the neural
networks weights and biases with many training
samples in a supervised learning approach. The network
fits correlations between the input data and the output labels, i.e.,
the two parameters $\tilde{p}_\mathrm{L}$ and
$\Delta \tilde{\rho}$.
This correlation can then be used to solve for input
never seen before.
The main difference to CSF is that
we do not directly adjust the parameters $\tilde{p}_\mathrm{L}$ and
$\Delta \tilde{\rho}$ for  each new shape separately,
but we rather adjust the weights
and biases of the neural network once by training with many shapes
and can then obtain parameters $\tilde{p}_\mathrm{L}$ and
$\Delta \tilde{\rho}$ almost instantly for any new shape without
further adjustment.
ML has recently been used to solve many
complex problems and is growing in popularity among
scientists, there is, to our best knowledge, no research showing the
capabilities of a ML approach for parameter extraction from shape
data of pendant drops. As a ML framework we use \emph{Keras}
\cite{chollet2015keras} and \emph{Tensorflow} \cite{tensorflow2015-whitepaper}.

\subsection{Architecture of the network}

The architecture of the neural net for pendant drop tensiometry is shown
in Fig.\ \ref{fig:network}.
The input to the network is essentially the same as the input to the numerical
fitting scheme - a discrete set of points along the  contour of a drop's
shape. For the ML approach we fix the number of sample points
along the shape to a specific sample count $d$, because the input shape
of a \emph{Dense}-Layer has to be of fixed size.
The resulting $d \times 2$ input matrix, consisting of the $\tilde{r}$- and
$\tilde{z}$-values of the $d$ sample points along the shape, then gets
flattened into a $2 d \times 1$ input vector. If the input data contains less
than $d$ samples the input vector gets zero padded and if the input data
contains more than $d$ samples the input is truncated while keeping the apex
coordinates. We use a sample count of $d = 226$ since an arc length step of
$10^{-2}$
between shape point samples gives shapes of class $2$ that generally
have less shape sample
points than $226$. Increasing the sample count $d$ increases
the complexity of the network and will slow the learning process.

The input vector is processed by a fully connected deep neural network with
\emph{Dense} neurons and \emph{Leaky-RELU} activation functions.
The \emph{Leaky-RELU} activation function aims to fix unwanted behaviour
occuring with regular \emph{RELU} activated neurons by replacing
the flat negative region of the \emph{RELU} function with a linear function
that has a finite slope $m \ll 1$ \cite{Xu2015}.

The first layer has an input dimension of $2d$ and outputs $512$ continuous
parameters, the second layer takes the $512$ outputs from the first layer and
processes them into $1024$ outputs which the third layer processes into $256$
outputs. The fourth layer has $16$ outputs and finally the fifth layer has $2$
output parameters, which are
the fitting parameters $\tilde{p}_\mathrm{L}$ and $\Delta\tilde{\rho}$.
The layer dimensions emerged from testing and show no overfitting with
the training data we use.

\subsection{Training of the network}

\begin{figure}
  \includegraphics[width=0.99\linewidth]{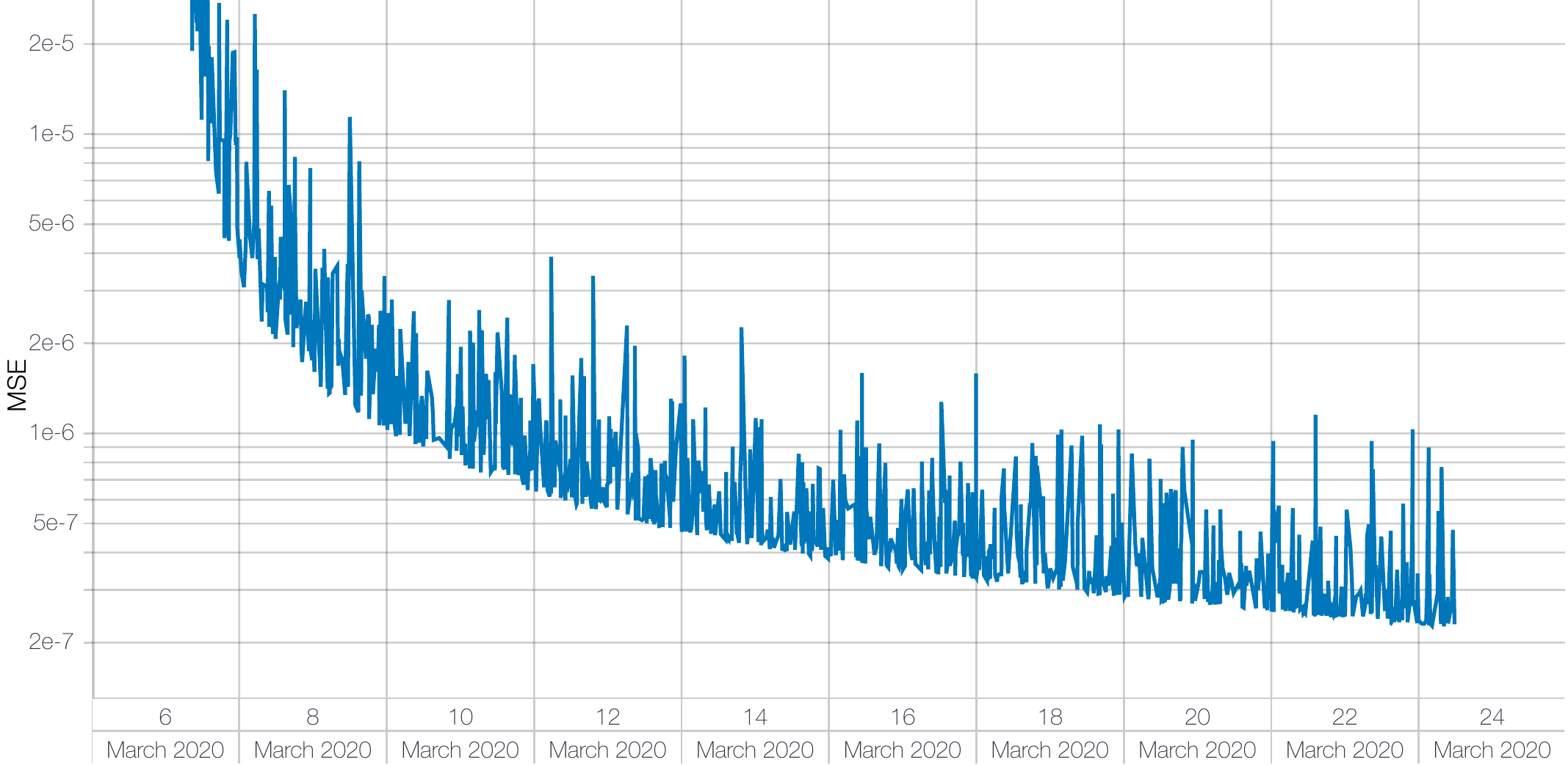}
  \caption{
    Evolution of the MSE  objective function (log-scale)
    with  training time.
  }
  \label{fig:training}
\end{figure}

\begin{figure*}
  \includegraphics[width=0.99\linewidth]{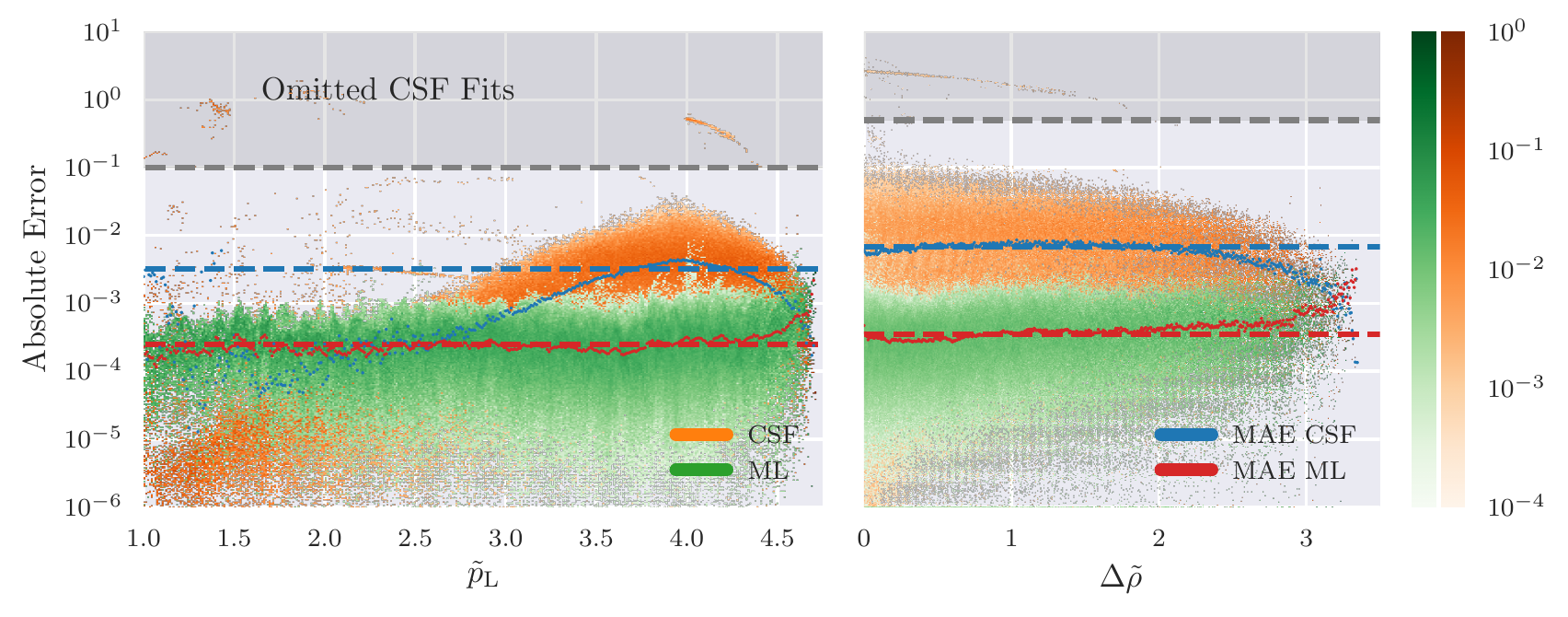}
   \caption{
     Comparison of the absolute errors for parameters
     $\tilde{p}_\mathrm{L}$ and $\Delta\tilde{\rho}$ between conventional
     shape fitting and machine learning for
     class $2$ shapes. The dashed
     lines  indicate the total mean absolute error (MAE) of all fits/guesses
     and labels.
     The relevant parameter
     for the determination of the interfacial tension $\gamma$
     is $\Delta \tilde{\rho}$.
   }
   \label{fig:comparison}
 \end{figure*}

 The  drop shapes for training  are generated using the
 numerical forward solution to the Young-Laplace problem
 with \emph{OpenCapsule}.
 At first, we select
  training shapes  randomly and uniformly selected from the
 relevant shaded ``triangle''
 enclosed by the yellow and blue bifurcation lines in the shape diagram
 Fig.\ \ref{fig:shapediagram} in the
 $\tilde{p}_\mathrm{L}$-$\Delta\tilde{\rho}$ plane, where
 shape classes $2$ and $3$ can exist.
 This choice of training set aims to
  obtain a neural network with  uniformly good performance
  in this whole parameter range. An alternative choice of training
 set will be discussed below.

As a performance
metric we pick the mean-square error (MSE) between the output guess and the
corresponding labels of the input data. An alternative error metric is the mean
absolute error (MAE), which does not penalize rare high amplitude errors
as much as MSE does.
We train the network with the MSE
$\frac{1}{2N}\sum_{n=1}^N
\left[ (\tilde{p}_\mathrm{L,in,n} -\tilde{p}_\mathrm{L,out,n} )^2 +
   (\Delta\tilde{\rho}_\mathrm{in,n} -\Delta\tilde{\rho}_\mathrm{out,n}
   )^2\right]$
of $N$ training shapes as an objective function.
The network is trained in batches of $N=100$  by backpropagation
using the \emph{Adadelta} \cite{Zeiler2012} gradient descent method
with an adaptive learning rate.

During the training period the network was trained for
approximately $90,000$ epochs, where one epoch consists of $0.5$ million drop
shapes and corresponding parameters $\tilde{p}_\mathrm{L,in n}$ and
 $\Delta\tilde{\rho}_\mathrm{in n}$.
On standard hardware (\emph{i3}-CPU with a \emph{GTX 970} GPU),
this training took approximately
3 weeks, see Fig.\ \ref{fig:training}.
The objective function is evaluated
with an independent set of $100.000$ shapes
between training epochs.
The final test set for the error comparison (see next section)
comprises another $0.9$ million shapes.

The network initially trains fairly quick,
as the precision
increases the learning rate decreases.
In total we see a steady sub-exponential learning process,
which we stop
at a precision of $\mathrm{MSE} = 2 \cdot 10^{-7}$, because the
precision gain per training time  diminishes.

\section{Results and comparison}
\label{sec:results}

The precision of the inverse solution by CSF
depends on how the accuracy
is set up in the numerics, more precision will
take more time to compute. The inverse fitting of the generated training data
with the CSF
algorithm takes between $0.25$ and $0.75$ seconds per shape to
compute on an \emph{i7}-CPU with $4.1\,\mathrm{GHz}$ with chosen settings of a
target parameter step of $10^{-2}$, i.e., an
absolute residual change of $10^{-2}$ in
$\tilde{p}_\mathrm{L}$ and $\Delta\tilde{\rho}$ during minimization
of the fitting error.

Once it is trained,
the neural network takes mere seconds to analyze all of the training
data images, only taking approximately $30$ microseconds per shape on a single
\emph{GTX 970} GPU. Further precision can be gained by
extending the learning process, changing the set of training shapes,
or changing the network's architecture.
We will explore the possibility of adapting the training
shape set below.

For the comparison of the fitting accuracy both
CSF and ML approach
are directly fed with  ``synthetic'' numerical  droplet shapes from
the output  of  the forward solution. This
creates a ``best case'' scenario for the  inverse solution.
Additionally, to calculate the performance of the CSF
implementation we only use those fits for which the
numerical inverse solution converged; including the shapes for which the
inverse algorithm failed, will worsen the mean error for the
numerical fitting. The ML approach is more robust and
has no problems
with failed inverse solutions: it generates a parameter guess for
any input shape.
We now want to compare the precision for both of these approaches.

First, we compare the actual parameters
$\tilde{p}_\mathrm{L}$ and $\Delta\tilde{\rho}$ of a given input shape
with the guesses from
the network and the results from the CSF by
their absolute errors in Fig.\ \ref{fig:comparison}.
We find  that the absolute errors of the
ML approach are roughly one order of magnitude lower
for both parameters on average
as intended by selecting training shapes uniformly from the relevant
parameter region.
The relevant parameter
for the determination of interfacial tension $\gamma$
is $\Delta \tilde{\rho}$, since
the dimensional parameters in its definition
\eqref{eqn:nondimensional_laplace} are commonly accessible in
experiment.
There are, however, phenomena in the errors
from the physics of pendant droplet shapes
   via the parameter sensitivity or insensitivity of these shapes.
The  inverse problem of determining the two fitting  parameters
can become  ill-conditioned  if
the shape becomes insensitive to changes in one of the parameters,
or it can become very well-conditioned if shapes are very sensitive.
We find that CSF performs exceptionally well
in the well-conditioned case, whereas ML outperforms
CSF in all other cases.

We observe in  Fig.\ \ref{fig:comparison}
that the determination of  the dimensionless
pressure $\tilde{p}_\mathrm{L}$ is generally unproblematic and
has a much smaller absolute error.
The reason is  that characteristic shape features such as the apex
curvature radius are uniquely determined by $\tilde{p}_\mathrm{L}$
via the Young-Laplace equation with good sensitivity.
CSF is producing smaller errors for low
$\tilde{p}_\mathrm{L}$ corresponding to larger  apex curvature radii, whereas
the ML approach has uniform errors, which is
due to the uniform selection of training shapes
from the  the first shaded ``triangle''
enclosed by the yellow and blue bifurcation lines in the shape diagram
Fig.\ \ref{fig:shapediagram}. As a result there are relatively
few shapes in the tip of the triangle corresponding to small
$\tilde{p}_\mathrm{L}$.
ML can reach
the performance of  CSF
by increasing the training set density in this region as we will
show below.


\subsection{The Worthington number as quality indicator}

\begin{figure}
   \includegraphics[width=0.99\linewidth]{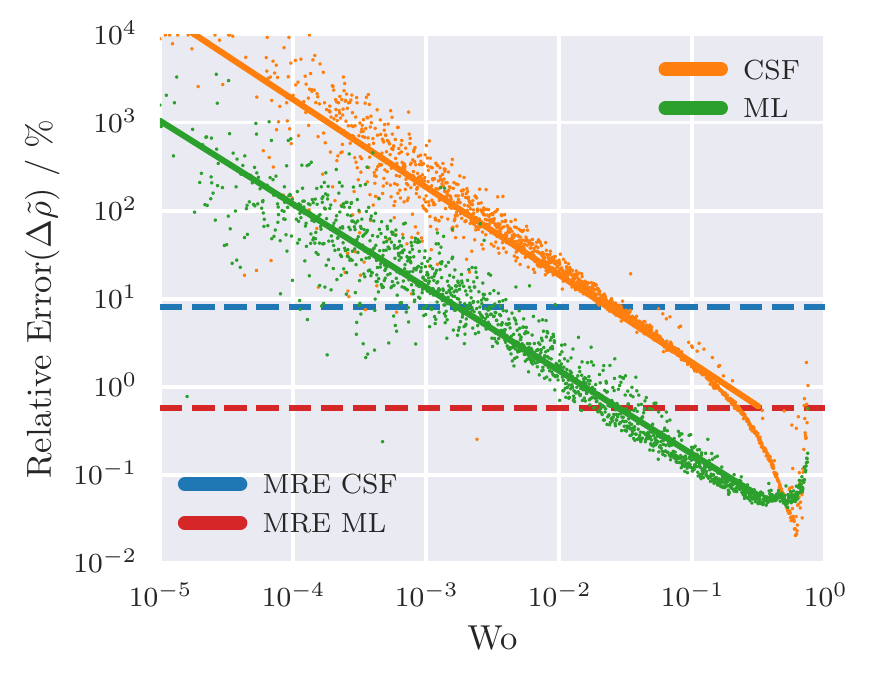}
   \caption{
     Scaling of the  total mean   relative error (MRE)
     of  $\Delta \tilde{\rho}$ with the Worthington number (\ref{eq:Wo}).
}
   \label{fig:Wo}
 \end{figure}

Figure \ref{fig:comparison} also shows that
 the absolute errors in the parameter  $\Delta \tilde{\rho}$
 are generally larger because the shape is less sensitive
 to changes in $\Delta \tilde{\rho}$.
 A uniform absolute error in $\Delta \tilde{\rho}$ will result
 in a relative error scaling as
 $\mathrm{MRE}_{\Delta \tilde{\rho}}\sim \Delta \tilde{\rho}^{-1}$.
 Deviations from this scaling point to particularly sensitive
 or insensitive shapes.
 In Ref.\ \citenum{Berry2015}, it has been proposed on a purely
 phenomenological basis that
the relative error in the fitting accuracy of $\gamma$,
which is proportional to the relative error in $\Delta \tilde{\rho}$,
is best described by
a parameter $ \mathrm{Wo}$ (Worthington number),
which is proportional to  $\Delta \tilde{\rho}$,
\begin{equation}
  \mathrm{Wo} \equiv \Delta \tilde{\rho} \frac{\tilde{V}}{\pi}
     = \frac{\Delta \rho g V}{\pi \gamma a}\,.
  \label{eq:Wo}
\end{equation}
This parameter measures the distance to the detachment volume according
to Tate's law (\ref{eq:Tate}) such that $\mathrm{Wo}< 1$ is bounded
and $\mathrm{Wo}\simeq 1$ corresponds
to a droplet close to detachment, while $\mathrm{Wo} \ll 1$ corresponds to
droplets far from detachment.
For a uniform absolute error in  $\Delta \tilde{\rho}$, we thus
expect a scaling  $\mathrm{MRE}_{\Delta \tilde{\rho}}\sim \mathrm{Wo}^{-1}$.
The scaling of the relative error  $\mathrm{MRE}_{\Delta \tilde{\rho}}$
for CSF and
ML approach is shown  in Fig.\ \ref{fig:comparison}.
 We find a power law scaling
 $\mathrm{MRE}_{\Delta \tilde{\rho}} =  a \mathrm{Wo}^\nu$
 and a fit in the linear region of the loglog plot
 in Fig.\ \ref{fig:comparison} gives the relative
 error scaling exponent and the scaling factor for
 CSF and ML,
\begin{align}
    \nu_\text{CSF} &= -1.00~,~  \nu_\text{ML} = -0.95
      \nonumber\\
     a_\text{CSF} &= -0.72~,~  a_\text{ML} = -1.71,
\end{align}
i.e., the exponent $\nu$ is indeed close to unity.

There is, however, the region of high Wo numbers
$\mathrm{Wo}>0.1$, where
CSF performs significantly better.
Focusing on this region, Berry {\it et al.} found an
exponent $\nu_\text{CSF} \approx -2$
indicating exceptionally small relative errors.
Based on the shape diagram Fig.\
\ref{fig:shapediagramV} we can actually rationalize this finding
and provide a theoretical basis for the use of the
Worthington number  $ \mathrm{Wo}$ as quality indicator in CSF.
Close to a bifurcation such as the bifurcation $1\to 1,2,3$, where
shapes  $\Omega =2,3$ appear (at the yellow lines), shapes
are most susceptible for parameter changes. This is evidenced, for example,
by the vertical tangent in the $\tilde{V}(\tilde{p}_\mathrm{L})$
relation in Fig.\ \ref{fig:pV} at this bifurcation.
Similarly, there is a vertical tangent
in the   $\tilde{V}(\Delta \tilde{\rho})$ relation.
Therefore, we expect exceptional shape sensitivity and, thus,
a very well-conditioned shape fitting problem in the vicinity
of this bifurcation.
We already pointed out  that this bifurcation happens almost at the
same volume as detachment for
$\Delta \tilde{\rho} \lesssim 1$, see shape diagram
Fig.\ \ref{fig:shapediagramV}.
Therefore, a parameter  $ \mathrm{Wo}\lesssim 1$ corresponds to
a regime close to the detachment volume and, thus, close to the
bifurcation  where shapes $\Omega=2,3$ appear and, therefore,
to the regime of a very well-conditioned shape fitting problem
(at least for $\Delta \tilde{\rho} \lesssim 1$).
Obviously, CSF
works very well in exactly such well-conditioned parameter  regions.
This rationalizes the use of the Worthington number
as quality indicator in CSF.
Interestingly, the critical
points at the tip of the  shaded ``triangles''
enclosed by the yellow and blue bifurcation lines in the shape diagram
Fig.\ \ref{fig:shapediagramV}, always lie
at $\mathrm{Wo}=1/2$ according to  the
bifurcation condition (\ref{eqn:bifurcation_cond}) and $\tilde{p}_\mathrm{cap}
= 2$. Therefore, all shapes above the blue dashed line
containing the critical points in the
shape diagram Fig.\
\ref{fig:shapediagramV} in the
$\Delta \tilde{\rho}$-$\tilde{V}$  parameter plane have
high Wo numbers $\mathrm{Wo}\ge  1/2$.
CSF should work very well and give the
best results in this region of the shape
diagram.
Figure  \ref{fig:comparison} shows that only in this region
CSF can outperform the ML approach.

ML gives
a uniform absolute error
over the full range of Wo numbers resulting in an exponent
$\nu_\text{ML} \approx -1$.
Therefore, the Worthington number Wo is less indicative for the
performance of the ML approach, at least
with the present set of training shapes uniformly distributed
in the
$\tilde{p}_\mathrm{L}$-$\Delta\tilde{\rho}$ plane.
For smaller Wo numbers,  ML
gives on average a full order of
magnitude more accurate estimates than
CSF, while being  four orders
of magnitude faster.

\subsection{Adapting the training of the network}

 \begin{figure}
   \includegraphics[width=0.99\linewidth]{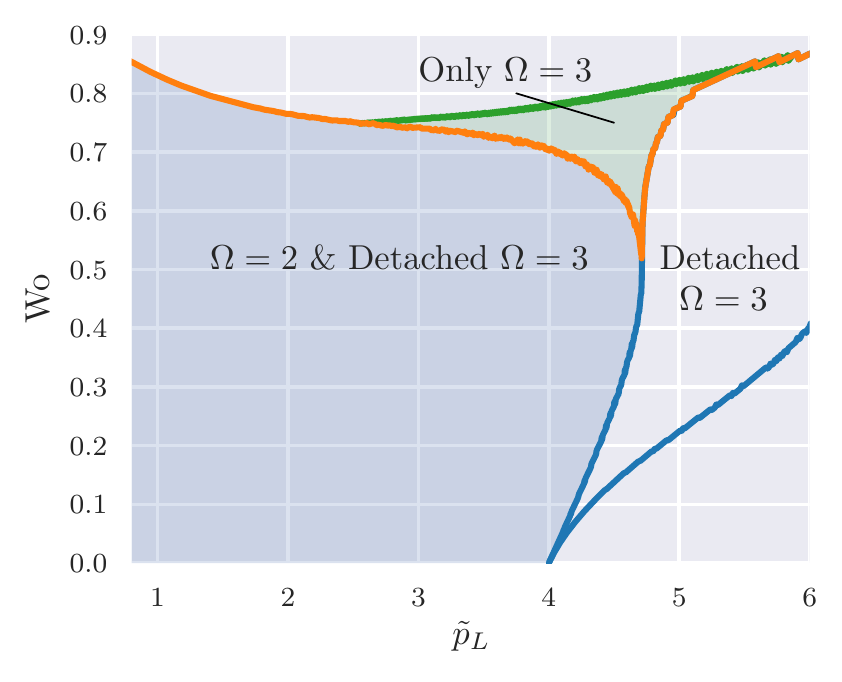}
   \caption{The training data sampling can be adjusted to be uniform in the
   $\tilde{p}_\mathrm{L}- \mathrm{Wo}$ plane. In the blue area only shapes of
   class 2 are sampled. In the green area only shapes of class 3 are sampled.
   We use the detachment condition discussed above to limit the sampling for
   class 3 shapes to pre-detachment shapes only.}
   \label{fig:Wo_p_L_sampling}
 \end{figure}

Finally, we want to try to improve the ML approach
such that it can handle all class 2 and 3 shapes while outperforming
CSF for {\it all} Wo numbers.
We can improve the performance of the ML approach
selectively in the regions of high Wo numbers by adapting our
training set such that it contains more shapes in this region.
Sampling the training set uniformly
from the shaded ``triangle''
enclosed by the yellow and blue bifurcation lines in the shape diagram
Fig.\ \ref{fig:shapediagram} in the
$\tilde{p}_\mathrm{L}$-$\Delta\tilde{\rho}$ plane
leads to a training set biased towards small Wo numbers.
Therefore, we adapt our training set such that it samples uniformly in the
$\tilde{p}_\mathrm{L}$-$\mathrm{Wo}$ plane depicted in Fig.\
\ref{fig:Wo_p_L_sampling}. In the new training set we include all
class 3 shapes
up to detachment (green sampling region in Fig. \ref{fig:Wo_p_L_sampling})
as well as all class 2 shapes (blue sampling region in
Fig.\ \ref{fig:Wo_p_L_sampling}).

Generating a set of shapes sampled uniformly in the
$\tilde{p}_\mathrm{L}$-$\mathrm{Wo}$ plane poses a problem,
since the relationship
between the control parameters of the simulation
$\tilde{p}_\mathrm{L}$ and $\Delta \tilde{\rho}$ and the sampling parameter
$\mathrm{Wo}$ is not known analytically. From data analysis we can extract
a phenomenological dependency between $\Delta \tilde{\rho}$, the sampling
parameters $\mathrm{Wo}$ and $\tilde{p}_\mathrm{L}$:
\begin{equation}
  \Delta \tilde{\rho} \left(\tilde{p}_\mathrm{L},\mathrm{Wo}\right) \approx
  \frac{\pi}{43} \mathrm{Wo}^{0.91}
  \tilde{p}_\mathrm{L}^{\pi \left( 1 - \frac{\mathrm{Wo}}{4}\right)} \,.
  \label{eqn:phenomenological_rho_Wo}
\end{equation}
This relation is based on an Ansatz $ \Delta \tilde{\rho}  \sim
\mathrm{Wo}^{1-\epsilon}  \tilde{p}_\mathrm{L}^{\delta}$
motivated by the definition of the Worthington number (\ref{eq:Wo}),
$ \Delta \tilde{\rho}  \sim \mathrm{Wo} \tilde{V}^{-1}$ and
an Ansatz $\tilde{V} \sim  \tilde{p}_\mathrm{L}^{-\delta}$ for the
pressure-volume relationship.
While \eqref{eqn:phenomenological_rho_Wo} provides a good mapping for
$\tilde{p}_\mathrm{L} < 3$ it lacks in accuracy for higher
$\tilde{p}_\mathrm{L}$, where it can not be used for the generation of an
evenly sampled training data set.

We ultimately generate the training set by the following algorithm,
which does not use the
relation (\ref{eqn:phenomenological_rho_Wo}).  First, we
pick $\mathrm{Wo}$ and $\tilde{p}_\mathrm{L}$ from a uniform distribution.
Second, we algorithmically search for the upper and lower boundary of the
shape diagram for $\Omega = 2$ at the picked $\tilde{p}_\mathrm{L}$.  Third,
we numerically calculate the upper boundary shape and from it the upper
boundary $\mathrm{Wo}^{\Omega = 2}_\mathrm{max}$. If the picked $\mathrm{Wo}$
is \emph{bigger} than $\mathrm{Wo}^{\Omega = 2}_\mathrm{max}$ we search for a
solution with $\Omega = 3$. Should the picked $\mathrm{Wo}$ be \emph{smaller}
than $\mathrm{Wo}^{\Omega = 2}_\mathrm{max}$ we search for a solution with
$\Omega = 2$.  Last, the search for the target $\mathrm{Wo}$ is achieved by
bisecting the interval between the upper and lower boundary of the
corresponding valid parts of the shape diagram Fig.\ \ref{fig:shapediagram}
in the $\tilde{p}_\mathrm{L}$-$\Delta\tilde{\rho}$ plane
and determining the corresponding value of  $\mathrm{Wo}$
till a numerically defined
precision for $\mathrm{Wo}$ is reached.

\begin{figure}
  \includegraphics[width=0.48\textwidth]{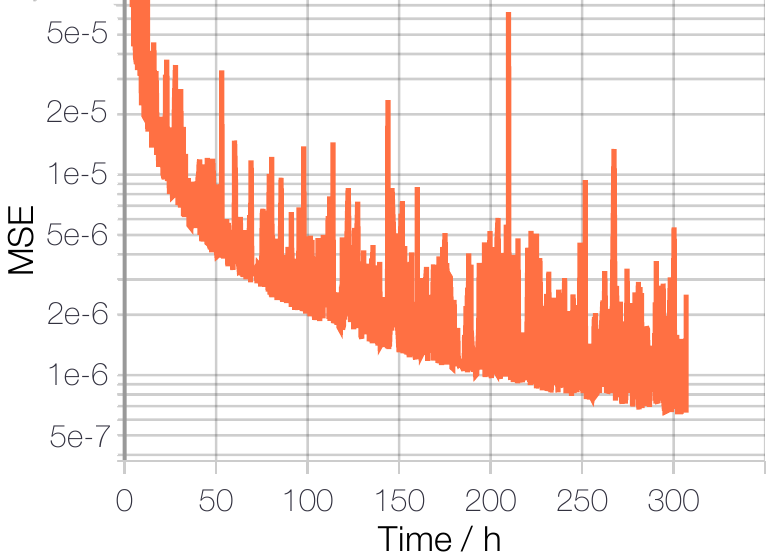}
  \caption{Training process of the adapted network
    in the first 300 hours of training.}
  \label{fig:training_Wo_sampling}
\end{figure}

This algorithm provides an evenly sampled training data set in the
$\tilde{p}_\mathrm{L}$-$\mathrm{Wo}$ plane, which we use to
train a new neural
network, for which the training process is shown in
Fig.\ \ref{fig:training_Wo_sampling}.

The new sampling produces shapes that have a longer total arclength $L$ on
average, thus we also modify the input sample count of the neural network to
be $d = 512$ and we append the pre-processed volume of the shape to the input
vector providing a new feature that could help reduce the complexity caused by
the increased sample count. Other pre-processed features available from the raw
shape data could also be provided to further increase accuracy while reducing
the complexity of the network.

\begin{figure}
  \includegraphics{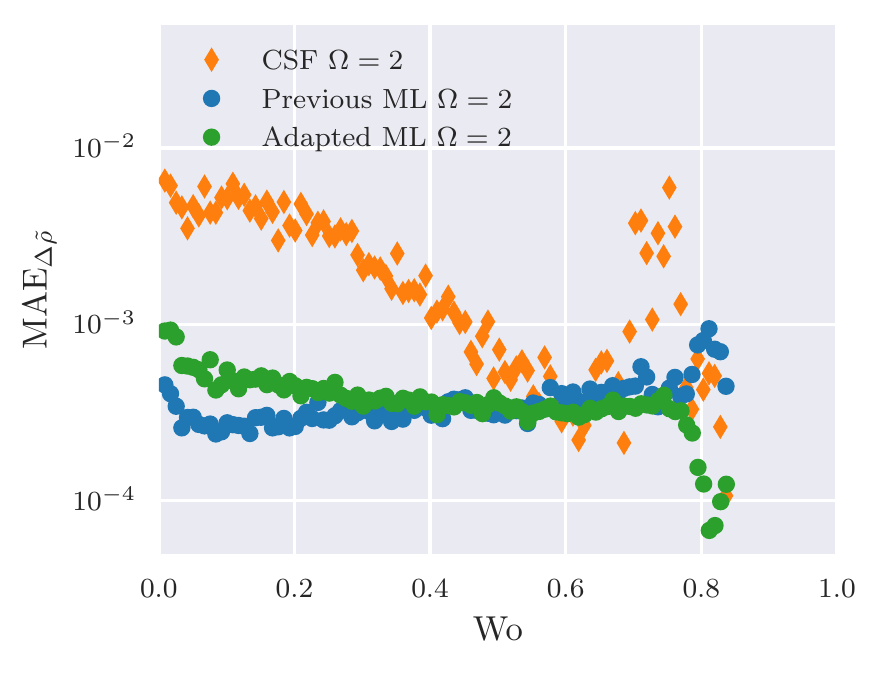}
  \caption{The absolute error of the adapted machine learning approach
    compared to the absolute error of the conventional shape fitting. Now
    class $2$ and $3$ solutions are used to train the network, however only
    class $2$ is shown.  We omit all CSF fits with an absolute error that is
    higher then $10^{-1}$, while considering \emph{all} ML predictions. }
  \label{fig:Wo_sampling_absoulte_error}
\end{figure}

The resulting network performs well over the full range $\mathrm{Wo} \in [0, 1]$
as can be seen in Fig. \ref{fig:Wo_sampling_absoulte_error}. The absolute error
is decreasing as $\mathrm{Wo}$ increases and the network gets
extremely accurate for
$\mathrm{Wo} \sim 0.8$. While the previous network performs better for small
$\mathrm{Wo}$, the adapted network is a full order of magnitude better than the
previous network for $\mathrm{Wo} \sim 0.8$.

The new network can also accurately solve
the inverse problem for class 3 solutions up to detachment.
In a comparison between CSF and the newly trained neural network for class 3
solutions we can see that the accuracy advantage of the CSF for high
$\mathrm{Wo}$ melts away by including class 3 solutions up to detachment.
This has to do with the fact that shapes become extraordinarily sensitive at the
bifurcation between shapes 2 and 3 but as the class 3 solutions approach the
detachment condition their $\mathrm{Wo}$ gets larger while the shape gets
increasingly \emph{insensitive} because class 3 shapes move \emph{away}
from the
bifurcation line while increasing $\mathrm{Wo}$ up to detachment.

The ML approach provides good accuracy for \emph{all} input shapes
and thus gives a more reliable predicted set of shape parameters.
The shape parameters predicted by the ML approach may also be
improved further
by using them as an initial guess in a CSF algorithm if needed.

\subsection{Noise tolerance comparison}

Comparing the ideal scenario of providing perfect input to both approaches
might give an insight into the capabilities of both approaches in a best case
scenario, it is, however, not realistic.
Any given solution technique has to work
with imperfect data in the real world. These imperfections might arise from a
limited camera resolution, an imperfect edge detection software or by
imperfections in the rest of the experimental setup. We want to discuss how
both approaches can handle noisy input data. For this we apply a Gaussian
blur to all shape coordinates $\vec{x}_i$ to transform them into a set of
distorted shape coordinates $\vec{x}'_i$. The Gaussian blur is centered around
the origin so its mean is given by $\mu = 0$ and the standard deviation $\sigma$
can be adjusted to create different noise amplitude scenarios.

\begin{figure*}
  \includegraphics[width=0.99\linewidth]{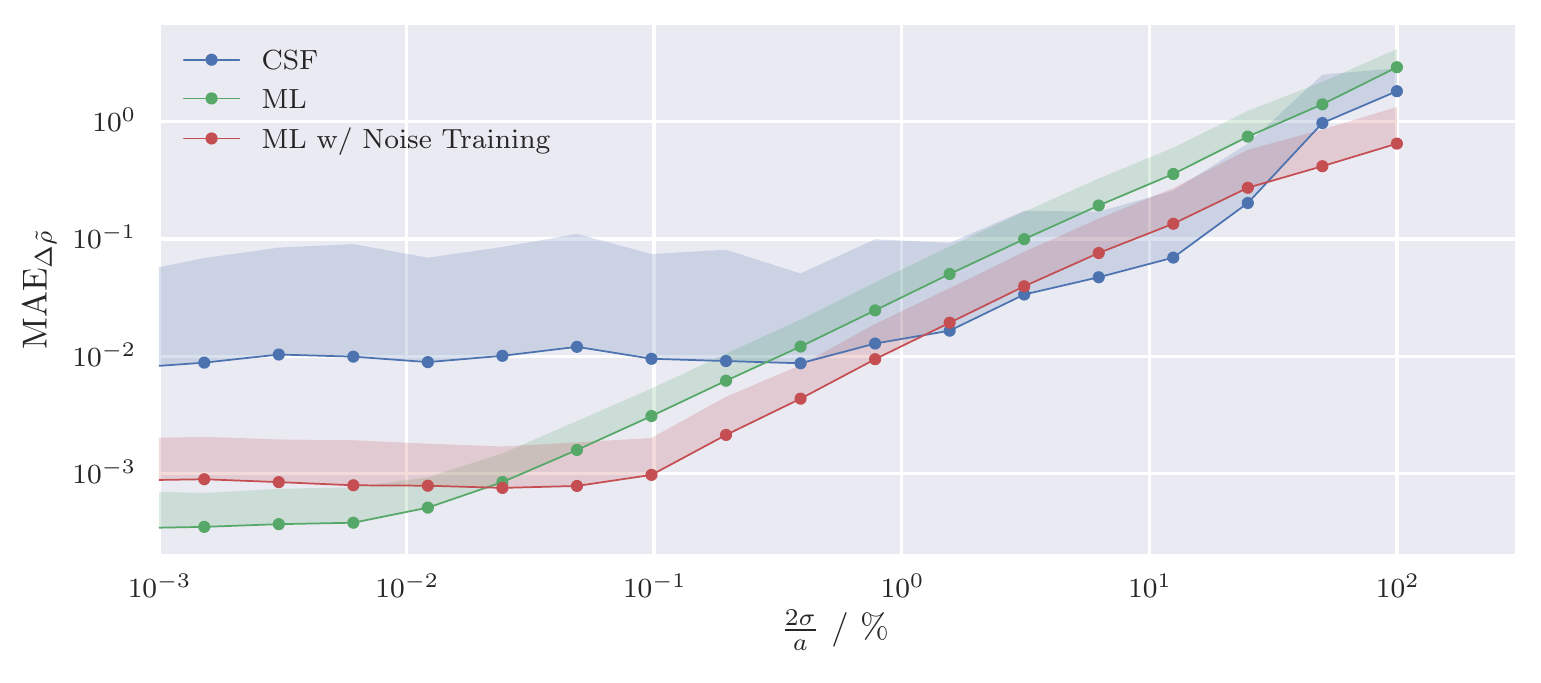}
  \caption{Comparison of the absolute error of class $2$ shapes with
  increasing Gaussian blur amplitudes added to the shape data. The shaded
  regions visualize the standard deviation of the mean absolute error. Because
  of the logarithmic scale only the upper bound is shown.}
  \label{fig:blur_comparison}
\end{figure*}

When comparing the performance of both methods in
Fig.\ \ref{fig:blur_comparison}
we can observe that the ML approach that has been trained on
undistorted data performs very well for low noise amplitude scenarios, but
is outperformed by CSF
for high noise amplitude scenarios.
Again, we can adapt the training set to improve the ML
approach.
To increase the real world performance of the ML approach we can
use noisy input data to train a new network that can handle noise better. We do
this by training on a set of approximately $500,000$ shapes with a Gaussian
blur applied to all shape coordinates, leading to a much improved noise
resistance as can be seen in Fig.\ \ref{fig:blur_comparison}, while also
maintaining almost equal precision in low noise amplitude scenarios. We also
notice that changing
from {\it{MSE}} to {\it{MAE}} as a training metric of the network improves
the accuracy for noisy data even further, because then large individual errors
do not dominate the overall mean error and the gradients in backpropagation
are less noisy, leading to improved precision.

\section{Discussion and conclusion}

We introduced a novel ML approach to
pendant drop tensiometry, where we train a deep neural network
with numerically generated training shapes (solutions of the
forward problem) for given pressure and interfacial tension
in order to solve the backward problem of interfacial tension
determination
for a measured (or synthetically generated) droplet shape.
We compare the performance of this mML approach to
CSF approaches to tensiometry.
The ML approach benefits from our ability to
generate a arbitrarily large set of droplet training shapes numerically by
solving the Young-Laplace equation (solving the forward problem)
and control the distribution
of training shapes in parameter space, which creates an
ideal setting for supervised deep learning.

In order to rationalize the structure of shapes
in parameter space
we first discussed the physics of pendant drops and
developed a simple classification
of  solution classes $\Omega=1,2,3,..$ by the number of bulges and necks.
We obtained shape diagrams as a function of dimensionless apex pressure
$\tilde{p}_\mathrm{L}$, dimensionless density difference
$\Delta \tilde{\rho}$ (which is also a measure of  capillary
diameter), and volume $\tilde{V}$, i.e., shape diagrams
in the $\tilde{p}_\mathrm{L}$-$\Delta \tilde{\rho}$ parameter plane
under pressure control
(Fig.\  \ref{fig:shapediagram}) and
in the $\Delta \tilde{\rho}$-$\tilde{V}$ parameter plane
under volume control (Fig.\  \ref{fig:shapediagramV}).
We identified the regions of existence of all shape
classes and their bifurcations within the
shape diagram. For pendant drops under volume control
the shape sequence  $\Omega=1\to 2\to 3$  is the
 sequence of energetically
 preferred states with shape $\Omega=2$ of a pendant
 drop with one bulge being the global
 energy minimum in a large volume range, i.e.,
 everywhere where it exists.
 We also identified the detachment line of maximal volume
 within the shape diagrams and obtained the
 bifurcation condition (\ref{eqn:bifurcation_cond}), which
 also gives an excellent description of the
 detachment volume.
 Based on the shape diagrams we can propose several
 training strategies for supervised learning in
 the ML approach.
 We start with  training shapes chosen uniformly in the
 $\tilde{p}_\mathrm{L}$-$\Delta \tilde{\rho}$ parameter plane.

The ML approach we
provide is novel and  performs well on this specific problem.
It is not
only more accurate than the tested conventional fitting
scheme in large parts of the parameter space, but it is also
orders of magnitude faster. Note that the precision of the
inverse solution by CSF
is bound by the precision target we provide and can
 outperform the precision of a neural network in the discussed ``best
 case'' fitting scenario in principle,
 but this will also take much longer. The hardware needed to
execute a once trained neural network are miniscule in comparison to
the hardware needed to perform numerical fitting of data sets in a reasonable
time.

We chose a standard accuracy for the CSF
approach and find that it outperforms the ML approach
only in the regime of high Worthington numbers Wo close to unity.
We can rationalize the use of the Worthington number as quality
measure in conventional fitting approaches based on the shape diagrams
by showing that high Wo numbers indicate a very well-conditioned
shape fitting problem, where shapes are sensitive to
parameter changes because they are closed to the shape bifurcation, where
shapes $\Omega =2,3$ appear.

This is the motivation to
 adapt the training set for the ML approach further
to contain shapes that are sampled with uniformly distributed
Wo number. Using this strategy the ML approach's
precision for high Wo numbers is increased further.

This improvement by adaptation of the training set shows that
there is certainly more potential for improvement in the ML
approach either via the choice of training set or by
further optimizing the network architecture which was a relatively
simple five layer deep network (see Fig.\ \ref{fig:network}).
Recurrent neural networks (RNN) could also be tested to further improve
performance and reliability, as well as convolutional neural networks for full
image input analysis. These network types are generally more demanding on the
hardware and could thus reduce the throughput of the network drastically. For
rheological problems that consider a series of images
- like a deflation experiment to determine the viscoelastic moduli - a long-
short-term-memory (LSTM) input layer can be used to process the time component
of the information in an efficient way to reduce the dimensionality of the
data for the attached fully connected part of the network, as we will show
in later work.
Further improvement to the fully connected network type we provided can always
be achieved by hyperparameter optimization and testing.

Because of the orders of magnitude faster computation time
the ML approach can also be used for high-throughput
analysis of droplet
shapes in a short amount of time, or - provided a fast pre-processing algorithm
- even real time video analysis in a dynamic experimental setting.
Further investigations into the capabilities of neural networks in
computationally taxing numerical
fitting procedures, like pendant capsule elastometry
\cite{Knoche2013, Hegemann2018} or even
viscoelastometry are the next step in future work, as the
conventional fitting approach for those
problems can be exponentially more demanding.

We make the deep neural network developed and trained within this work
publicly available via GitHub \cite{GitHubRepo} for further
use in pendant drop tensiometry.






\bibliography{paper}

\end{document}